# Superconducting Accelerator Magnets

*S. Sanfilippo*
Paul Scherrer Institute, 5232 Villigen, Switzerland

**Abstract**
This course introduces key aspects of superconducting magnet technology in accelerators: basic principles, superconducting materials (NbTi, $Nb_3Sn$, ReBCO), wire and cable architectures, and fabrication methods. Compared to copper or permanent magnets, superconducting systems require cryogenics and complex protection schemes but enable superior performance. Core challenges—like flux pinning, magnetization effects, quench behavior, mechanical forces interception, power tests and magnetic measurements—are addressed through examples of magnets from PSI and CERN.

## 1 Introduction

In the context of accelerating climate change, reducing electrical energy consumption and developing compact, energy-efficient, and sustainable magnets for particle accelerators has become a strategic priority. Superconducting magnet technology offers a promising pathway towards more environmentally responsible research infrastructures, enabling higher magnetic fields with drastically lower operational losses compared to conventional resistive systems. This course provides a short overview of the superconducting technologies used in particle accelerators, focusing on their physical principles, material properties, design methodologies, manufacturing techniques, and operational challenges. A classification of low- and high-temperature superconductors (LTS and HTS) is presented, covering materials such as Nobium-Titanium (NbTi) Nobium-Tin, ($Nb_3Sn$), and Rare Earth Barium Copper Oxyde (Cuprate Superconductor, ReBCO). Their respective domains of application are introduced in terms of critical current density, field limits, and thermal stability. The course explores cable technologies and manufacturing strategies such as "wind-and-react" for brittle conductors. Superconducting magnets are defined as systems based on cryogenically cooled superconducting coils. Compared to normal-conducting copper or permanent magnet designs, they offer access to higher fields, reduced energy consumption, and novel integration possibilities, but also pose significant design constraints in terms of quench protection, cryogenics, and mechanical reinforcement. Material selection is covered with emphasis on flux pinning, filament structure, and magnetization effects. The course covers all the aspects to manufacture a superconducting magnet—from conductor processing and coil winding to impregnation, structural assembly, and cryogenic testing. Key challenges such as quench training, field quality perturbations, and protection systems (CLIQ, heaters, dump resistors) are discussed through case studies from major institutes like CERN and PSI. The course concludes with an overview of the decay and snap-back phenomena of the magnetization, dynamic effects caused by persistent currents in superconducting strands and cables.

## 2 Superconductivity: overview of the main characteristics

### 2.1 No resistance and Meissner effect

This paragraph offers a brief overview of the fundamental properties of superconductivity, particularly as they apply to accelerator magnet technology. This overview is not intended as a substitute for in-depth study. To support deeper understanding, a selection of recommended books (Bibliography) is provided at the end of the course, covering the physics of superconductivity, material science, and its application in accelerator magnets. Superconductivity is a quantum phenomenon observed in certain materials when cooled below a critical temperature Tc. In this state, two essential features emerge: zero electrical resistance (R=0, for T<Tc) and the Meissner effect, which is the total expulsion of magnetic fields from the interior of the material—reflecting perfect diamagnetism (Fig. 1). These properties allow superconductors to carry current without energy loss, making them ideal for generating strong and stable magnetic fields. The Meissner effect is especially important: unlike a perfect conductor that merely traps magnetic fields during cooling, a superconductor actively expels them upon entering the superconducting state, thanks to persistent shielding currents flowing at the surface. This results in a magnetization M=−H, as shown in M(H) diagram. The transition between the normal and superconducting state is bounded by three critical parameters which are the most important parameter of merits: the critical temperature Tc, the upper critical field Bc2 and the critical current density Jc. These parameters are linked together and define a critical surface in the (T, B, J) space, within which the material remains superconducting (Fig. 2).

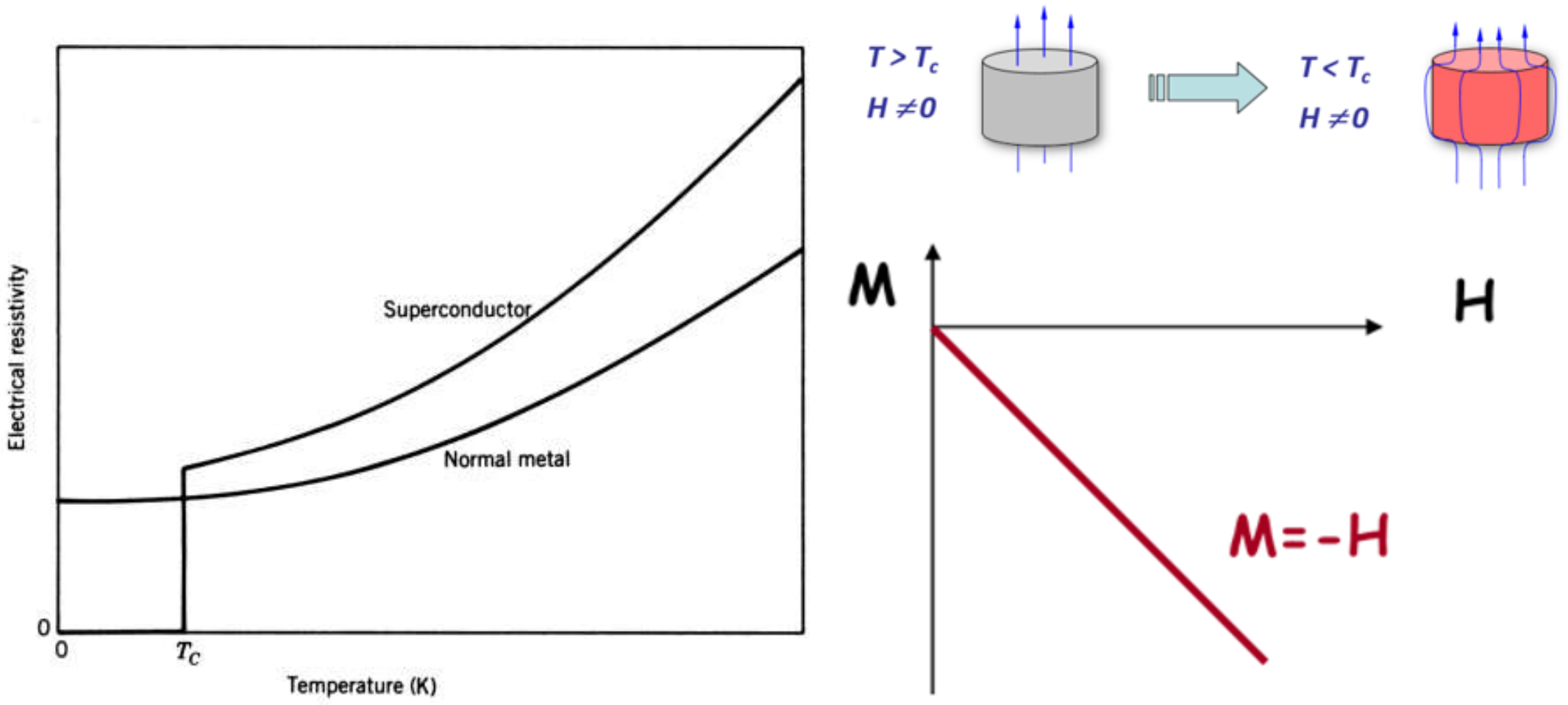

**Fig. 1:** Superconductors for T<Tc are perfect conductors (R=0, left) and display perfect diamagnetism (c= -1-, Right)

Superconductors are broadly categorized into low-temperature superconductors (LTS), such as NbTi and $Nb_3Sn$, and high-temperature superconductors (HTS), like YBCO and Bi2212. HTS materials have a $T_c$ above 30 K and can operate at higher temperatures and magnetic fields, but all technical superconductors require cryogenic systems (e.g., liquid helium or nitrogen) to remain in the superconducting state.

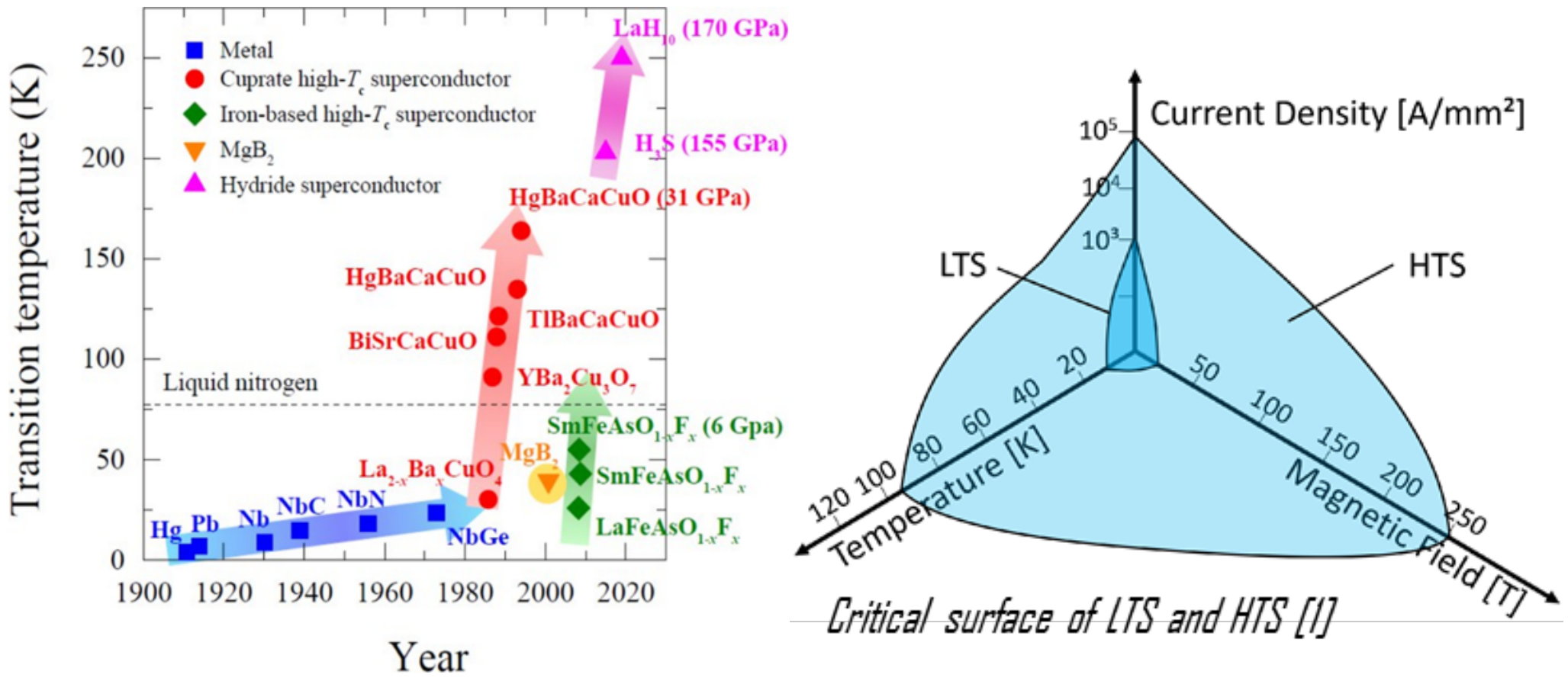

**Fig. 2:** Left -Classification of superconductors along their critical temperature and years of discovery. Right- critical volume or the superconducting state bounded by the critical temperature $T_c$ (J, B), the critical current density $J_c$ (T, B) and the upper critical field $Bc_2$ (T). The picture shows the reduced domain for the LTS superconductors ($T_c$ <30 K) with respect to the HTS ones (Pictures from [20] and [33]).

## 2.2 Type I and type II superconductors

Superconductors are broadly classified into Type I and Type II, based on how they respond to applied magnetic fields.

- Type I superconductors exhibit a complete Meissner effect: they expel all magnetic fields from their interior as long as the external magnetic field H remains below a critical value $H_c$. Once H exceeds $H_c$, superconductivity is destroyed, and the material enters a normal (non-superconducting) state. The magnetization curve for Type I materials is thus sharp and linear up to $H_c$, after which it drops to zero. Examples are pure elemental superconductors like mercury or lead.

- Type II superconductors exhibit a more complex behavior with a mixed state. For fields below a lower critical value $Hc_1$, they behave like Type I materials, expelling all magnetic flux. But between $Hc_1$ and an upper critical field $Hc_2$, they enter a mixed state where magnetic flux partially penetrates the material in the form of quantized vortices (also called fluxoids). These are modelized as cylindrical regions where the material is locally non-superconducting containing a quantum of magnetic flux $\Phi_0=h/2$. The rest of the material remains superconducting, and shielding supercurrents and transport currents coexist. The vortices can move under Lorentz forces, leading to dissipation unless they are immobilized by defects. For this reason, microstructural engineering to introduce pinning centers is essential to improve the performance. Above $Hc_2$ the material transitions fully to the normal state.

In type II superconductors mixed state, these vortices tend to arrange themselves into a hexagonal lattice (for low $T_C$ superconductors) to minimize repulsive interactions. The ability to support this intermediate state makes Type II superconductors much more robust under high magnetic fields, with upper critical fields $Hc_2$ reaching up to several tens of Tesla (e.g., 20–100 T), compared to milli Tesla-to-Tesla ranges for Type I materials.

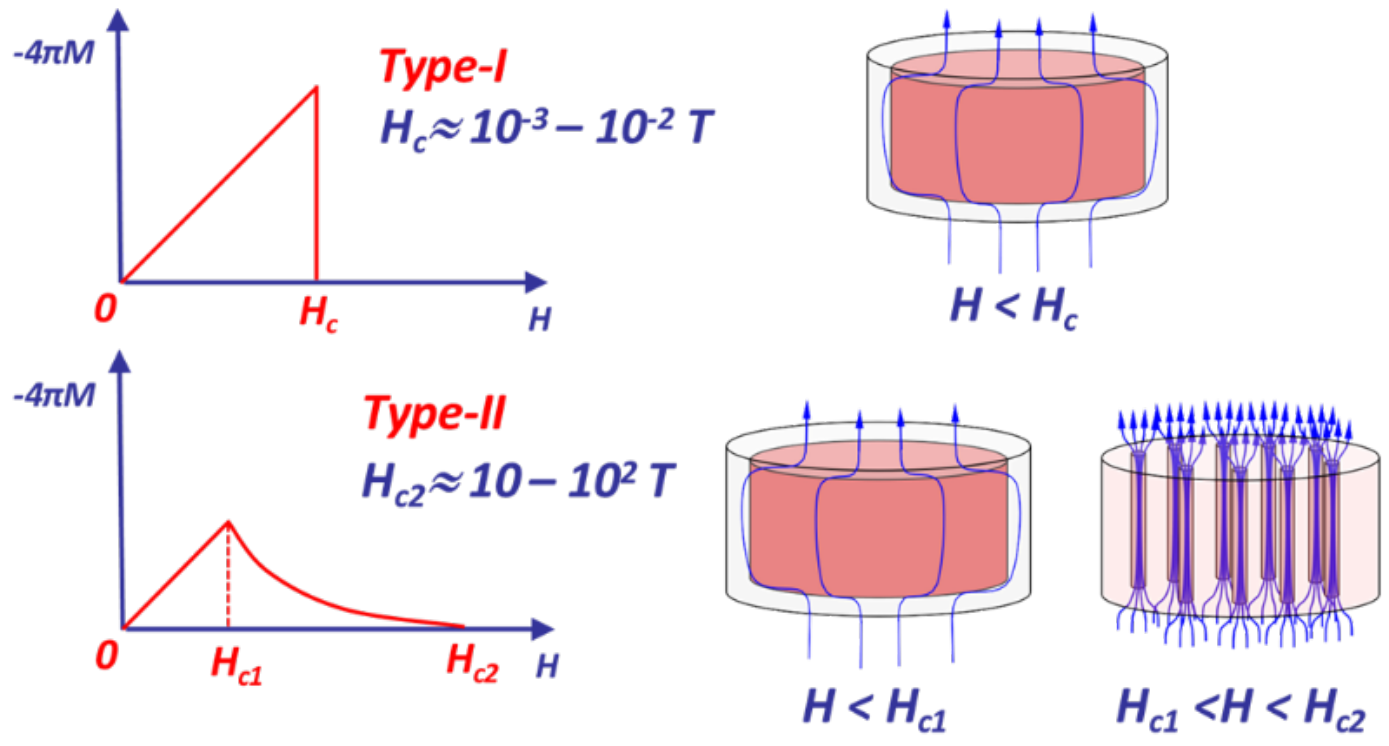

**Fig. 3:** Magnetization variation with respect to temperature for Type 1 and Type 2 superconductors (Pictures from the bibliographic reference [Senatore 2020]).

## 2.3 Vortex pinning in LTS and HTS superconductors

In Type II superconductors, the transport current $J_T$ flows through the material in the presence of a magnetic field in the form of quantized vortices. These vortices consist of a normal (non-superconducting) core surrounded by circulating superconducting currents. As the current flows, the vortices experience a Lorentz force given by $\mathbf{F_L}=\mathbf{J_T}\times\mathbf{B}$ which tends to set them in motion leading to flux flow—a form of energy dissipation that introduces electrical resistance. This directly undermines the superconducting state, as it reintroduces losses into what should be a zero-resistance material. To prevent this, practical superconductors rely on the pinning of the vortices: the immobilization of vortices by microstructural defects such as inclusions, dislocations, grain boundaries, or engineered nano-defects. These defects act as pinning centers that "trap" the vortices. As long as the pinning force $F_p$ exceeds the Lorentz force, the vortices remain stationary, and no energy is dissipated. Effective pinning occurs when the size of the defect is comparable to the coherence length ξ—the typical size of a vortex core. For low-temperature superconductors (LTS), ξ~4–15 nm, while for high-temperature superconductors (HTS), ξ~1–2 nm. Thus, the manufacturing process must be carefully optimized to introduce the right type and distribution of defects that enhance pinning without degrading other properties.

In summary, vortex pinning is essential for enabling practical superconductivity. It allows superconductors to carry high currents in strong magnetic fields without energy loss. This is particularly crucial for accelerator magnets, where both current density and field strength are extreme. Without pinning, superconductivity would be quickly destroyed by vortex motion and the associated dissipation. Vortex pinning in superconductors is achieved by introducing defects in the material that disrupt the regular superconducting structure and act as anchors for magnetic vortices. These defects can be classified based on their dimensionality and origin (Fig. 4):

### *2.3.1 Point defects (0D)*

These are zero-dimensional defects such as vacancies, grain boundaries, and precipitates. They can occur naturally or be artificially introduced using techniques like:

- Melt Texturing, which creates secondary-phase inclusions (e.g., green phase).
- Heavy-ion irradiation, generating columnar (1D) tracks but starting from point-like collisions.
- Chemical doping (e.g., with zirconium) to introduce nanoscale inhomogeneities.
- Substrate modifications, such as depositing nano-particles, which produce spherical pinning centers.

Point defects are highly effective when their size is comparable to the coherence length ξ of the vortex core. This results in strong core pinning, which maximizes the pinning force $\mathbf{F_p}$ and thereby enhances the critical current density $\mathbf{J_c}$ and the irreversibility field $\mathbf{B_{irr}}$**.** This field corresponds to a transformation of the nature of the vortex lattice: below $\mathbf{B_{irr}}$ the vortices are pinned, while above they move freely ("vortex lattice melting"), leading to vanishing hysteresis and the onset of resistivity [5].

### *2.3.2 Extended defects (1D and 2D)*

Extended defects include dislocations, twin boundaries, and columnar tracks, which extend over one or two dimensions. While not as efficient for core pinning as point defects, they are essential—particularly in high-temperature superconductors (HTS) operating above 4.2 K—because they reduce vortex motion caused by thermal fluctuations, which is more prominent at elevated temperatures.

### *2.3.3 Intrinsic pinning*

Some materials, such as cuprate superconductors, exhibit intrinsic pinning due to their layered structure, where superconductivity is confined to $CuO_2$ planes. This naturally restricts vortex motion along specific directions and contributes to overall pinning performance.

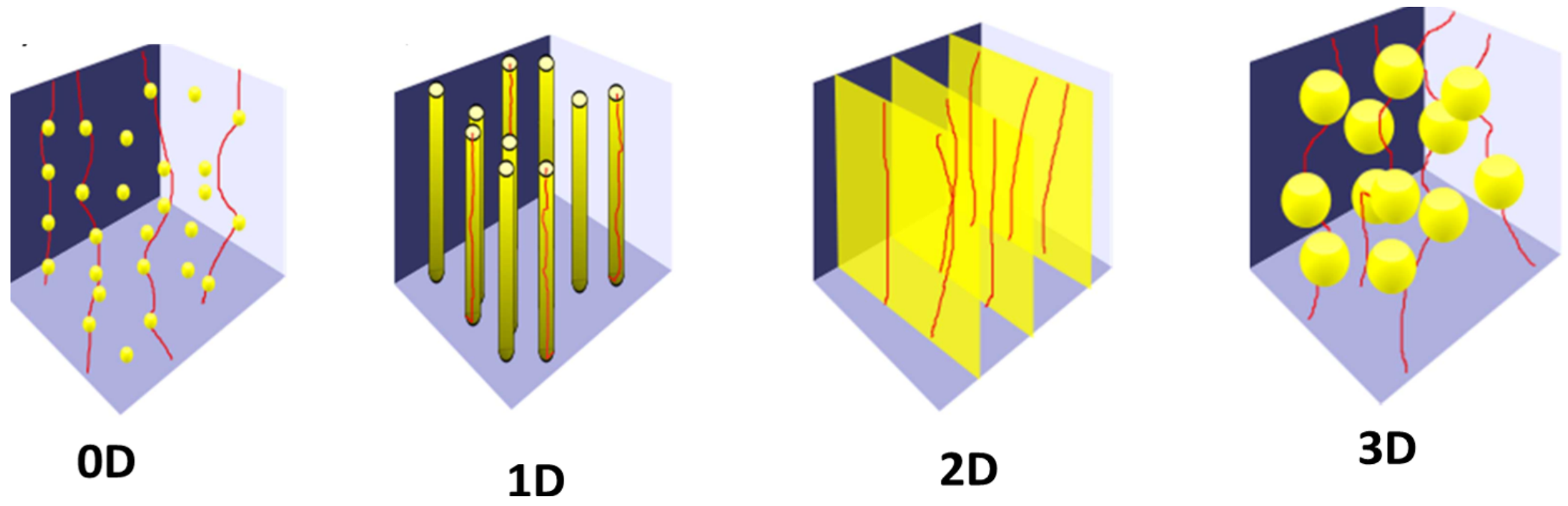

**Fig. 4:** Various pinning centers: Point defects (0D), linear defects (1D), planar ones (2D) and 3D artificial pinning centers [22].

## 2.4 Magnetic phase diagram of types II superconductors

Type II superconductors exhibit a complex magnetic phase diagram due to the interplay of various physical phenomena influencing the nature and the dynamic of the vortices.

### *2.4.1 Meissner State ($H<Hc_1$)*

Below the lower critical field $Hc_1$, the superconductor completely expels magnetic fields from its volume. This is the ideal superconducting state, exhibiting the Meissner effect, where the magnetic induction B=0 inside the material, and resistance is zero. In this region, there are no vortices.

### *2.4.2 Mixed State ($Hc_1<H<Hc_2$)*

When the applied field exceeds $Hc_1$, magnetic flux begins to penetrate the superconductor, forming a lattice of vortices (also called fluxoids). Each vortex has a normal-conducting core where superconductivity is suppressed, surrounded by circulating superconducting currents. This is known as the mixed state or Shubnikov phase.

Within this region, the behavior of the superconductor depends on three competing energy scales:

- Pinning energy $U_{pinning:}$ Associated with defects in the material that trap vortices and prevent their motion.
- Elastic energy $U_{elastic}$ : Comes from vortex-vortex interactions. Vortices repel each other and tend to arrange themselves in a hexagonal Abrikosov lattice, minimizing interaction energy.
- Thermal energy $k_BT$: Tends to depin vortices, making them mobile and leading to energy dissipation (flux flow), which undermines superconductivity.

### *2.4.3 Normal state*

At the upper critical field $Hc_2$, the superconducting state disappears and the material enters the normal resistive state. The entire volume is penetrated by magnetic flux, and electrical resistance reappears.

### *2.4.4 Behavior Across the Mixed State*

- At low fields just above $Hc_1$ vortex spacing is large, and the lattice rigidity dominates. Vortex-vortex interactions control their distribution.
- At intermediate fields, the pinning force **Fp** becomes essential. It determines the maximum current density the superconductor can carry without vortex motion. The force must counteract the Lorentz force $\mathbf{F_L}$ on the vortices.
- At higher fields approaching $Hc_2$ the number of vortices becomes so high that pinning centers are saturated, and vortex motion becomes increasingly difficult to suppress.

### *2.4.5 Case of the HTS*

HTS materials such as YBCO and Bi2212 operate at much higher temperatures (typically 20–50 K) and exhibit significantly different vortex dynamics due to:

- higher operating temperatures → larger thermal energy $k_BT$,
- smaller coherence lengths $\xi$ → weaker pinning forces,
- higher anisotropy → reduced vortex stiffness (lower $U_{elastic}$).

This means that in HTS, thermal fluctuations are much more important than in LTS. These fluctuations can cause the vortex lattice to melt, creating a vortex liquid state where vortices are highly mobile even below $Hc_2$ . The practical limit of superconducting material is set by the irreversibility line $B_{irr}(T)$ which lies well below Bc2(T), in particular for superconductors with high crystallographic anisotropy (the Bi2223). Above Birr, vortices become unpinned due to thermal activation and starts to move. In this region the conductor shows finite resistance, even though it is technically still in the superconducting state. This drastically limits the useful operating region for HTS, particularly at high temperatures T>50K, making effective pinning strategies and cooling essential. The plots of Fig. 5 highlight the magnetic phase diagram of HTS superconductors and the experimental determination of the $Bc_2$ and $B_{irr}$ lines for particle superconductors. The review article from G. Blatter et al. provides an extensive description of the pinning and vortices states in HTS superconductors [5].

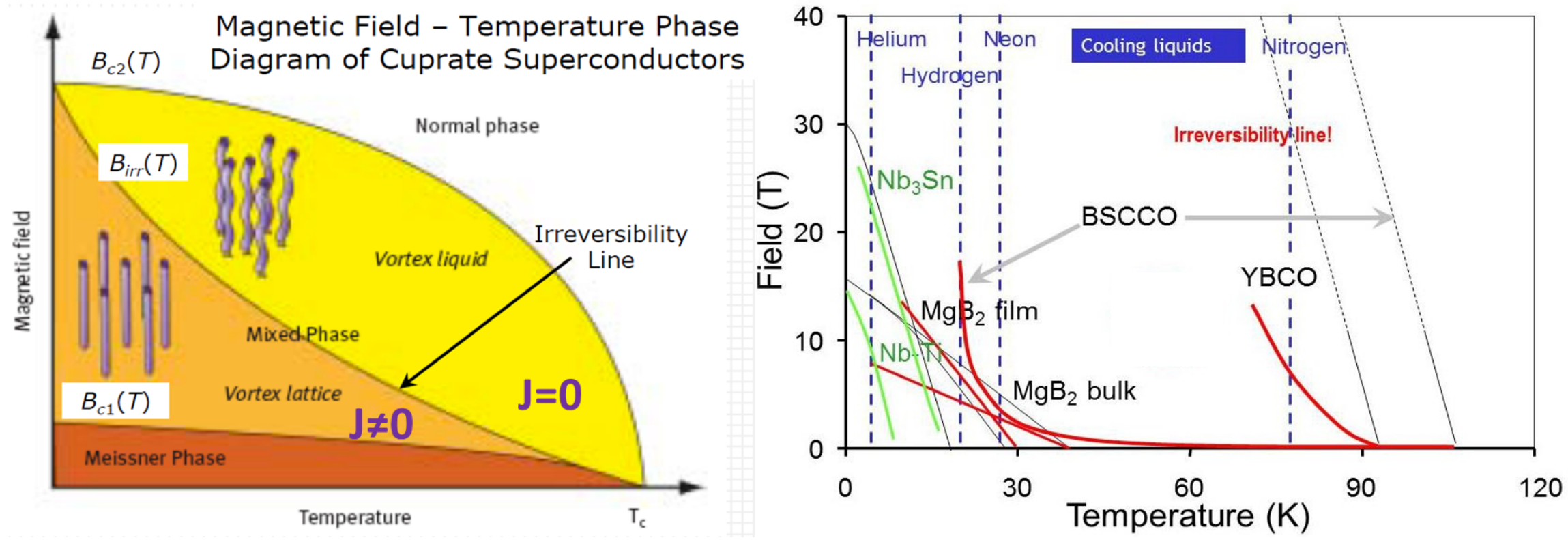

**Fig. 5:** Left - Magnetic phase diagram for HTS superconductors; Right – Upper critical field and irreversibility lines for LTS and HTS practical superconductors [30].

To conclude this overview, we are considering the case of practical superconductors. A practical superconductor must combine high performance with stability under operational conditions. Key figures of merit include a high critical current density ($J_c$), large upper critical field ($Bc_2$), and high critical temperature ($T_c$) to reduce cryogenic demands. For accelerator applications, strong and optimized vortex pinning is essential to prevent flux motion and energy dissipation. Pinning centers must match the coherence length (ξ) and be homogeneously distributed. In high-temperature superconductors (HTS), thermal fluctuations must be controlled to extend the domain below the irreversibility line ($B_{irr}$).

## 2.5 Loss of the superconducting state - quench of a magnet

A quench is a sudden and irreversible transition of a superconducting material to its normal (resistive) state. It marks a breakdown of superconductivity due to one or several external or internal disturbances (mechanical, electromagnetic or thermal perturbations such as winding pack motion, AC losses, loss of vacuum or coolant, nuclear heat loads...). When a quench occurs, the affected region of the superconductor loses its zero-resistance property and becomes resistive, generating heat that can propagate rapidly through the magnet. If not controlled the consequences of a quench could be severe for the superconducting coil and magnet:

- voltage increases across the superconducting coil, a voltage below 1 kV has to be limited,
- generation of thermal and electromagnetic forces,
- cryogen expulsion due to local boiling (e.g. helium evaporation),
- risk of damage to the magnet or associated systems.

There are two main types of quenches based on the mechanism that leads to the transition as illustrated in Fig. 6.

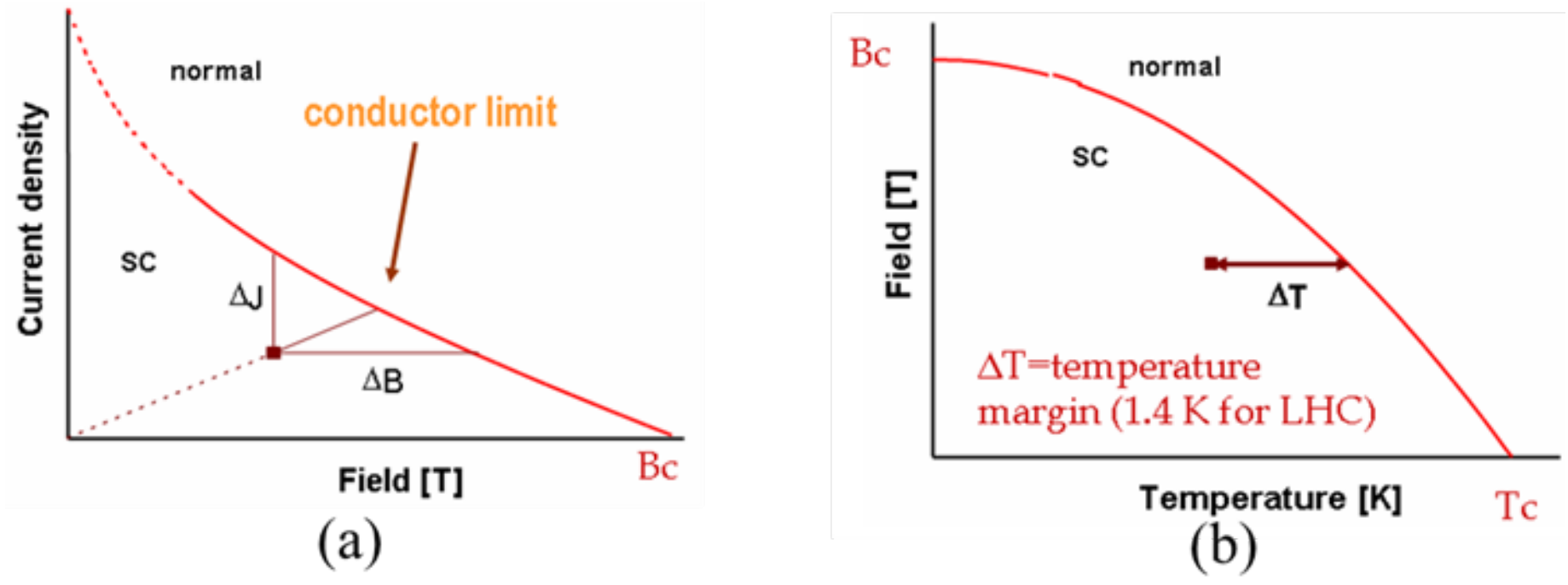

**Fig. 6:** (a)-the critical surface J(B) is crossed by a local increase of the current or the magnetic field. (b)-the critical surface B (T) is crossed by a local increase of the temperature [7].

### *2.5.1 Conduction-Limited Quench*

This type occurs when the critical surface $J_c$ (B,T) is exceeded due to:

- an increase in current,
- and/or an increase in magnetic field B.

The current-carrying capacity of a superconducting conductor depends on the local magnetic field and temperature. When either parameter increases beyond the limit, the current density J exceeds the critical current density $J_c$ causing the material to become resistive.

### *2.5.2 Energy deposited quench*

This quench occurs due to a local increase in temperature, which reduces the material's capacity to remain superconducting under the same current and magnetic field. A small heat input can be enough to raise the local temperature T beyond the critical temperature $T_c$ especially when the operating margin is small.

This type is dominant in many practical cases, particularly:

- at high field and current,
- with complex mechanical assemblies like cables and coils.

Thermal stability can be studied in terms of Minimal Quench Energy (MQE) also called energy margin, generally defined as the uniform energy density deposited in the conductor to initiate the quench propagation without recovery of superconducting state. The temperature margin and moreover the energy margin must guarantee to balance potential energy deposition for the largest spectrum as possible of perturbation events. Stability margins in superconducting magnets are critical for preventing accidental quenches due to thermal disturbances. Stability refers to a magnet's ability to operate safely without losing superconductivity when exposed to small energy inputs. To ensure such stability, engineers define several types of operating margins. One key type is the load line margin, which measures how far below the critical current density Jc the magnet current operates. It is defined as $1-J_{op}/J_c$, $J_{op}$ is the operating current density. This margin is visualized in the right graph Fig. 7, showing the magnet's position below the critical surface $J_c$ (B), providing a buffer against current-induced quenching.

Another important margin is the temperature margin, defined as the difference between the current sharing temperature $T_{cs}$ and the actual operating temperature $T_{op}$. $T_{cs}$ is the temperature at which the superconducting cable begins to share current between the superconductor and its stabilizing normal metal (typically copper or aluminum) when carrying a given transport current. This parameter indicates how much temperature increase the system can tolerate before losing its superconducting properties. The left graph Fig. 7 illustrates this by showing the intersection between $J_{op}$ and the temperature-dependent critical surface $J_c$(T). A larger temperature margin ensures that the magnet can absorb more heat without transitioning into a resistive state, which is particularly important for systems subject to time-dependent or spatially varying heat loads. If a local temperature increases ΔT due to a local source of heat ΔQ, exceeds the margin, the quench condition is met. If $\Delta T > T_{cs} - T_{op}$, the quench occurs.

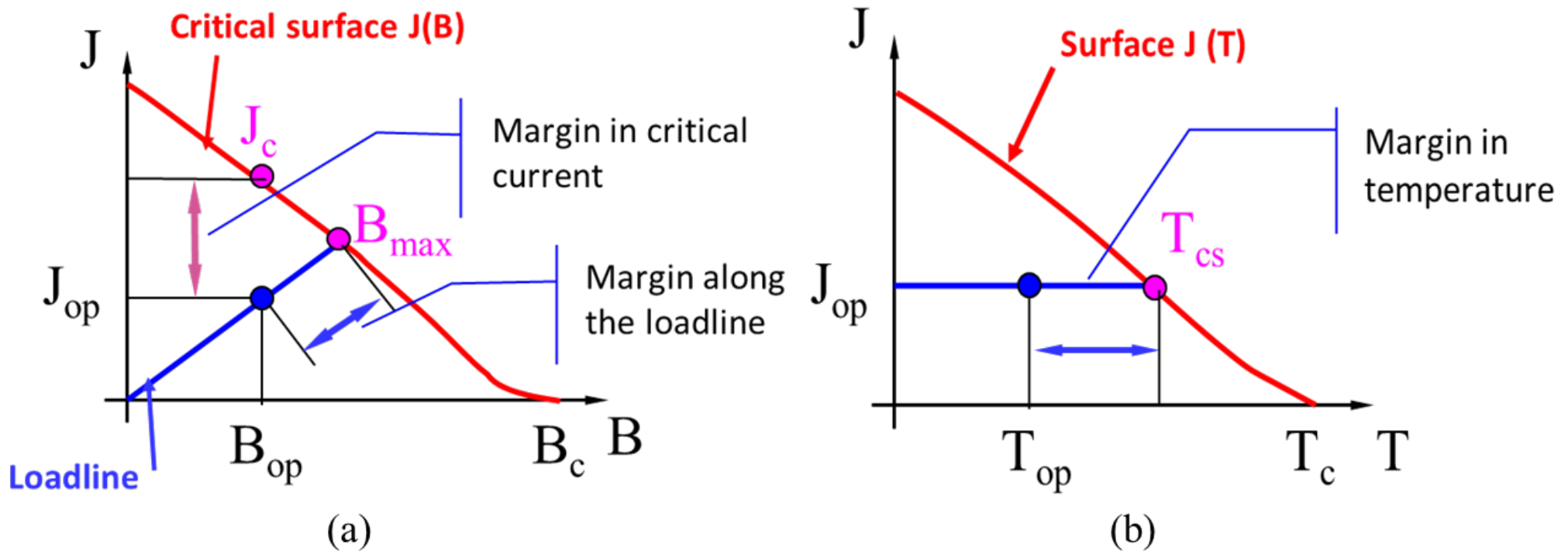

**Fig. 7:** (a)-Margin in critical current. (b)-Margin in temperature for a safe operation [7].

Finally, the critical field margin considers the magnetic field environment. It is typically expressed as the ratio $B_{op}/B_{max}$, where $B_{op}$ is the operating field and $B_{max}$ is the maximum field the conductor can withstand at the given operating current.  In practice, NbTi-based superconducting magnets are often operated with typical margins: about 50% of the critical current, 75% of the critical field, and 1–2 K of temperature margin. These values offer a practical balance between performance and stability, ensuring that the magnet remains in the superconducting state under normal operational fluctuations.

Possible source of quenches are:

- Conductor limited, when the conductor is carrying a current beyond its critical current density, initiating the loss of superconductivity.
- Mechanical disturbances such as cable strand motion, micro-slips, or coil deformation and vibration can release local energy and initiate a transition to the normal state.
- Intrinsic sources like conductor instability (flux jumps), insulation damage, broken strands and AC losses can lower the stability margin and lead to a quench.
- Cryogen flow interruption, due to pump failure or blockage, reduces heat removal capacity. Even small regions with insufficient cooling can undergo local boiling of helium and initiate a thermal runaway.
- Thermal sources like excess heating in splices or current leads may act as initiation points for a quench. Other sources are external heat deposition like beam deposition induced particle showers or nuclear heat, which can deposit energy in the conductor, locally raising the temperature above the critical limit.

The graph in Fig. 8 illustrates the energy-time spectrum of different disturbances, from fast flux jumps and wire motion to slower processes like heat leaks and AC losses. All of these disturbances deposit energy into the system, increasing the local temperature, and if this energy is not absorbed by the conductor's enthalpy, a quench occurs.

$$\Delta Q_{quench} = \int_{T_{bath}=1.9K}^{T_c(B,J)} c_{eff}(T)\,dT$$

This simple equation reflects the enthalpy margin—the energy needed to heat the conductor from its operating temperature up to the point of quench. If a disturbance injects more energy than $\Delta Q_{quench}$, the conductor's temperature will rise above Tc, causing a quench. Therefore, increasing the heat capacity, operating at lower temperatures, or maximizing the critical temperature (via material choice or reducing $B_{op}$ and $J_{op}$) will enhance the quench stability.

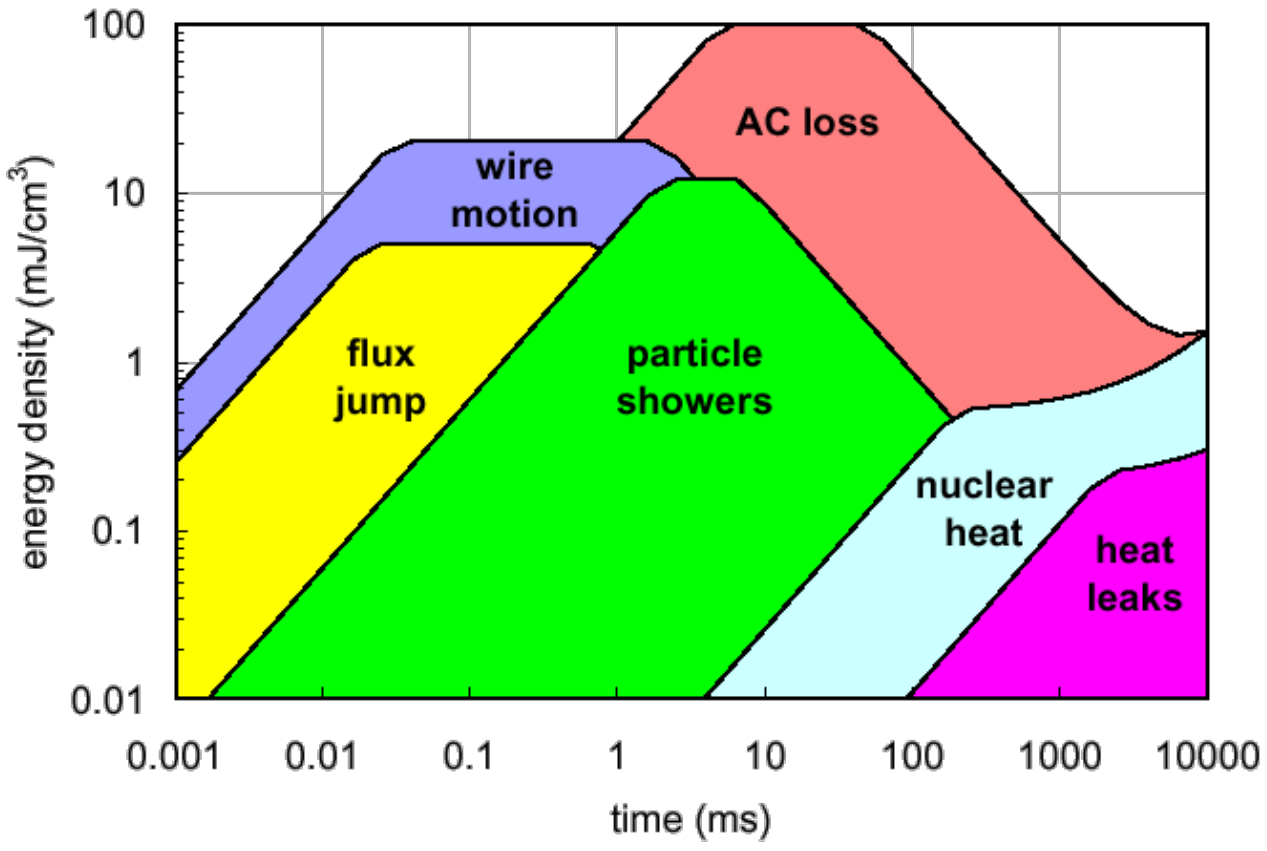

**Fig. 8:** Disturbance spectra of accelerator magnets [11]

## 3 Practical superconducting materials

Practical superconductors are engineered to meet the demanding requirements of high-field/high current applications by being manufactured in versatile and robust forms. They are produced as multifilamentary strands, containing thousands of superconducting filaments embedded in a stabilizing matrix, typically copper. This configuration enhances both current sharing and thermal stability, helping to avoid quench. In addition to strands, superconductors are fabricated as wires, tapes, multi-tapes, and cables to accommodate different geometries and performance needs. Cable formats include Rutherford cables, Roebel bars, CORC® cables, and CIC (Cable-in-Conduit) designs, each optimized for specific current capacity, flexibility, and mechanical strength. These formats allow superconductors like NbTi, $Nb_3Sn$, $MgB_2$, ReBCO, and Bi2212 to be used effectively in accelerator magnets, MRI machines, fusion reactors, and power applications. We will now overview the key steps in the fabrication of commonly used low-temperature superconductors (LTS)—namely NbTi and $Nb_3Sn$—as well as high-temperature superconductors (HTS) such as ReBCO. Each conductor type requires specific processing techniques tailored to its material properties and application requirements.

### 3.1 NbTi wires

The fabrication of NbTi wires, one of the most widely used practical superconductors (notably for LHC magnets), involves a multi-step industrial process optimized for mechanical flexibility, scalability, and reasonable superconducting performance. The NbTi alloy (typically containing 46–48% Ti by weight) is ductile, making it ideal for wire drawing.

Key Steps in NbTi wire fabrication are listed in the following [10]:

1. Extrusion: A composite billet, with a niobium-titanium core surrounded by a copper matrix, is sealed and heated, then extruded through a die to form a long, continuous hexagonal rod.
2. Stacking: Several of these extruded rods are stacked together inside a larger copper tube to create a multifilamentary billet, essential for forming wires with many fine filaments.
3. Heat Treatment: The stacked assembly undergoes thermal cycles to produce α-Ti precipitates, which serve as flux pinning centers, enhancing the wire's critical current performance.
4. Cold Drawing: The wire is drawn through successive dies, reducing its diameter and elongating it. This step improves mechanical strength and electrical uniformity and reduces the final filament size.

5. Twisting and Insulation: The drawn wires are twisted to reduce AC losses and then coated with insulation to prepare them for magnet winding. Final steps include spooling, quality testing, and annealing if needed.

This process produces flexible, multifilamentary superconducting wires with good mechanical and electrical properties, optimized for low and high-field, cryogenic applications.

### 3.2 $Nb_3Sn$ wires [10]

$Nb_3Sn$ is a key low-temperature superconductor for high-field magnets (up to 15 T). For an operating temperature of 4.2 K, the critical temperature $T_c$ is around 18 K and the upper critical field $B_{c2}$ is of 23–26 T, which is relatively high compared to NbTi. Unlike NbTi, $Nb_3Sn$ is brittle and highly sensitive to strain. The value of $J_c$ is dependent of this additional parameter. Its superconducting phase (A15 $Nb_3Sn$) is not present in the precursor wire but is formed through a high-temperature "reaction" process (typically 660°C for ~180 h). Because of its brittleness, this heat treatment must be performed after winding the magnet coil—an approach known as "wind-and-react". As shown in Fig. 9a, the current-carrying capacity of $Nb_3Sn$ is strongly influenced by applied mechanical strain. At low strain levels, the reduction in performance is reversible, but beyond a certain threshold, irreversible degradation occurs due to microstructural damage. This strain sensitivity imposes strict mechanical constraints in magnet design and highlights the importance of optimizing both the conductor architecture and the mechanical support structure to preserve performance under operational stress. The $Nb_3Sn$ fabrication steps involve assembling multifilament billets using Nb and Sn precursors, extrusion-drawing of the wire, and forming cables and coils prior to heat treatment. The fine grain size (150–200 nm) in the reacted $Nb_3Sn$ is essential for efficient vortex pinning. The performance of $Nb_3Sn$ strongly depends on strain: at low mechanical strain, the superconducting current can recover reversibly, but beyond a critical threshold, irreversible microstructural damage causes degradation in performance. This makes mechanical reinforcement and strain management critical in $Nb_3Sn$ magnet design. Three main fabrication routes exist (Fig. 9b): (1) the bronze process using a CuSn alloy as Sn source offers fine filaments and high strength but lower $J_c$; (2) internal tin (Sn rod source) gives very high $J_c$ with larger filaments; and (3) powder-in-tube (Sn from $Nb_3Sn$ powders) yields the highest $J_c$ but with less filament refinement.

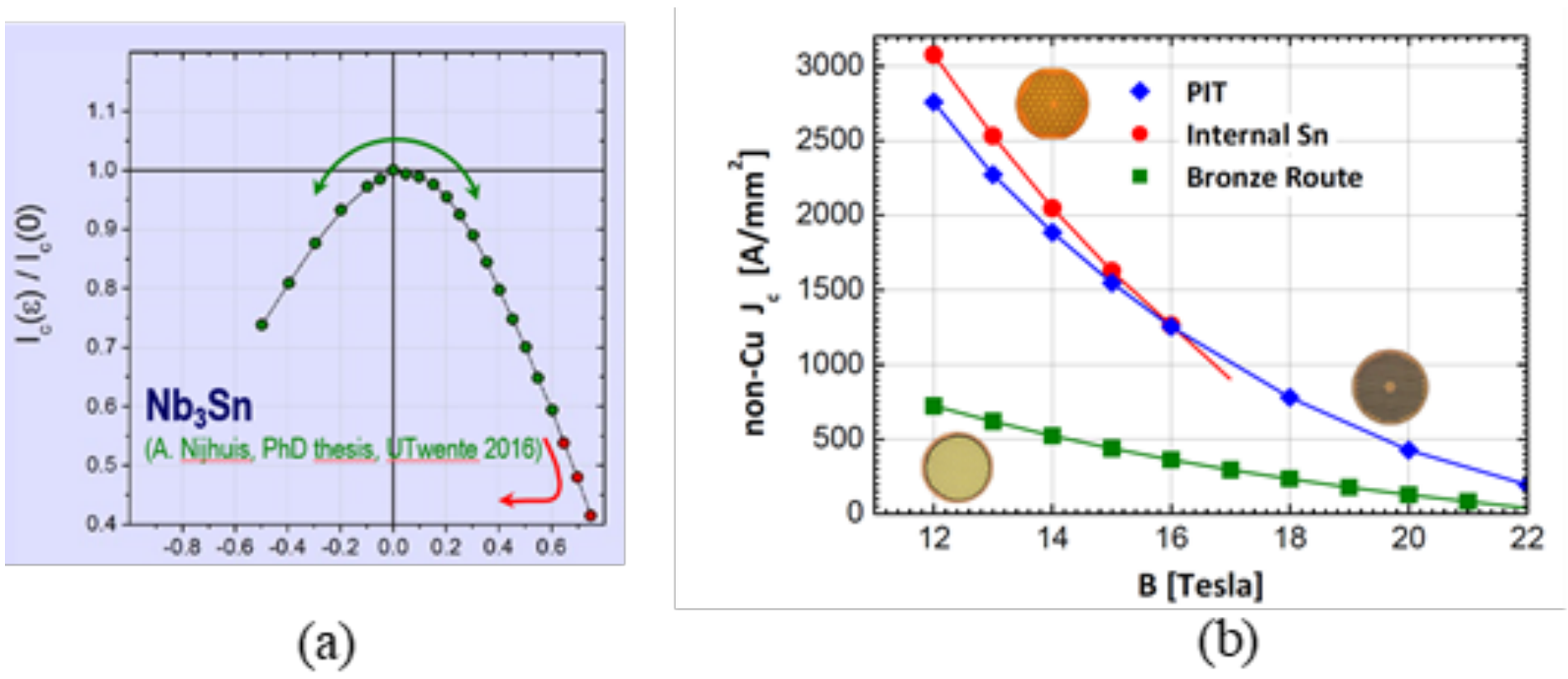

**Fig. 9:** (a) Strain dependence of the normalized critical current density [18]; (b) Variation of the critical current density with the applied magnetic field as a function of the production route [Senatore 2020].

### 3.3 LTS conductor: from filaments to cables.

#### *3.3.1 Multifilamentary conductors*

In low-temperature superconductors (LTS) such as NbTi and $Nb_3Sn$, the wires used in magnets are fabricated as multifilamentary composites: thousands of fine superconducting filaments are embedded in a copper matrix. This structure is crucial for optimizing performance in practical applications.

The diagram in Fig. 10 illustrates the origin of magnetization and hysteresis in a superconducting filament due to vortex pinning and shielding currents. In a Type II superconductor, magnetic flux penetrates the material in quantized vortices. Persistent shielding currents (in red) circulate near the surface of each superconducting filament, opposing changes in magnetic field and leading to a net magnetization M. This magnetization is not linear but exhibits a hysteresis loop due to the pinning of vortices — the magnetic response depends on the field history. The width of the hysteresis loop ΔM is proportional to the critical current density $J_c$ and the effective filament diameter d. Larger filaments result in stronger shielding currents and thus larger hysteresis, leading to higher AC losses and potential field quality perturbations in magnets. Therefore, controlling M, $J_c$ and d is essential for optimizing the performance of superconducting wires in accelerator and magnet applications.

To reduce these effects, the filament diameter is minimized—typically between to 6 and 50 µm (HERA14 µm, LHC 6-7 µm, HL-LHC 50 µm, FCC target filament 20 µm). This decreases the magnetization and hysteretic losses while maintaining the same critical current density $J_c$. The size limit of the filaments to avoid flux jumps are related to the critical temperature $T_c$, the density g and the specific heat C of the material by the relation:

$$d \leq \frac{2}{J_c}\sqrt{\left(\frac{3\gamma\, C\,(Tc - T)}{\mu_0}\right)}\;.$$

For conductor design, higher $J_c$ and operation near $T_c$ require smaller filament diameters for stability, while materials with high specific heat and density permit larger filament sizes. By keeping the filament size d below this limit, the heat generated can be absorbed by the filament's heat capacity without raising its temperature above $T_c$.

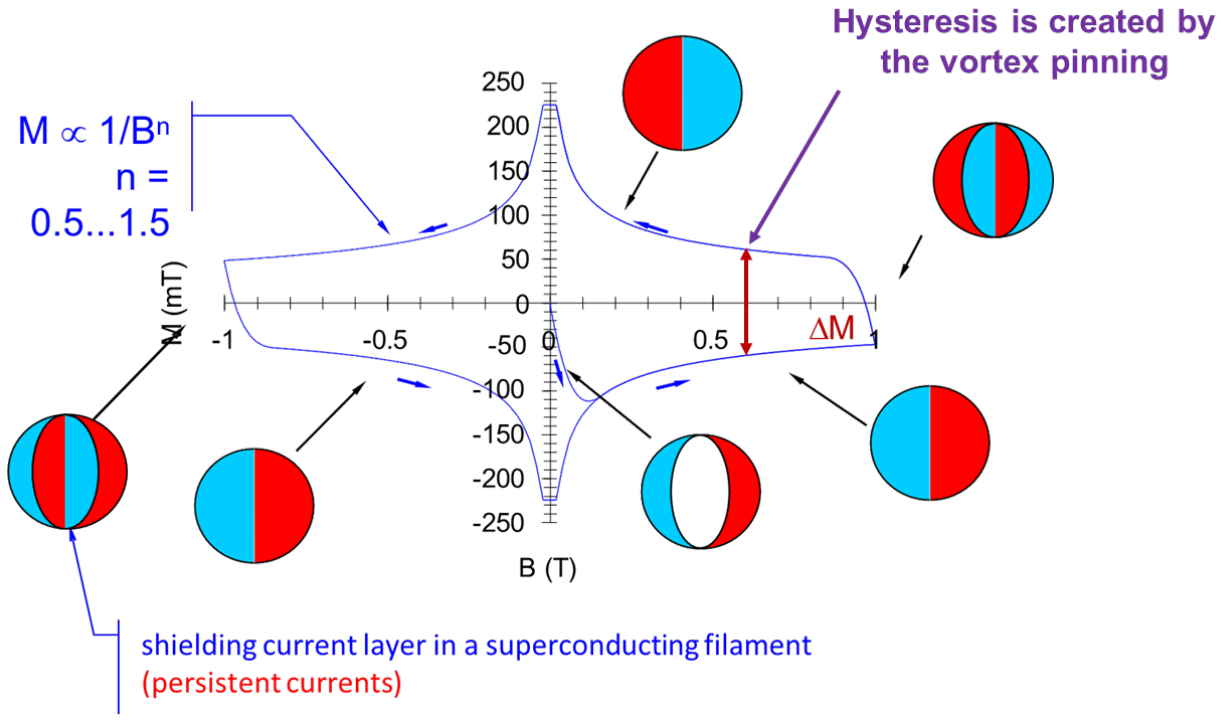

**Fig. 10:** Magnetization inside a filament with applied field created by the persistent current shielding the magnetic field. The hysteresis is created by the pinning of the vortices [24].

A trade-off has however to be found: if the filaments are too small and too closely spaced without sufficient electrical decoupling, inter-filament coupling currents may appear under time-varying fields. This creates large loops of circulating current and increases the effective filament diameter $d_{eff}$, leading again to significant losses and reduced field quality. This undesirable coupling is mitigated by twisting the filaments with a short pitch (e.g., 12–30 mm) before the final drawing step.

### *3.3.2 Stabilizer*

In Low-Temperature Superconductors like NbTi, the superconducting filaments are embedded in a matrix of normal metal called the stabilizer (Fig. 11), usually high-purity copper (Cu) or sometimes aluminum (Al). The stabilizer is essential for ensuring the wire's electrical, thermal, and mechanical stability. Stabilizer matrix plays the following roles:

- Improve the current carrying capacity: The stabilizer provides a parallel path for current when the superconductor is partially resistive (e.g., near the current sharing temperature $T_{cs}$). This improves the wire's total current-carrying capability and provides a safety margin. Practically, all current flows through the superconductor for $T \leq T_{cs} < T_c$. The current begins to share between the superconductor and the stabilizer while $T_{cs} < T \leq T_c$ and for T exceeding $T_c$, superconductivity is lost. All the current flows in the stabilizer. If cooling is insufficient, this leads to a quench, an irreversible and potentially damaging transition.
- Enhance Mechanical Properties: It improves the mechanical robustness of the wire, helping to withstand mechanical stresses from winding, thermal contraction, and Lorentz forces in high-field applications.
- Improve the quench Protection and Heat Spreading: During a quench (loss of superconductivity), the stabilizer:
  - conducts the current to prevent damage,
  - spreads and dissipates heat to reduce local overheating,
  - delays or avoids irreversible transition.

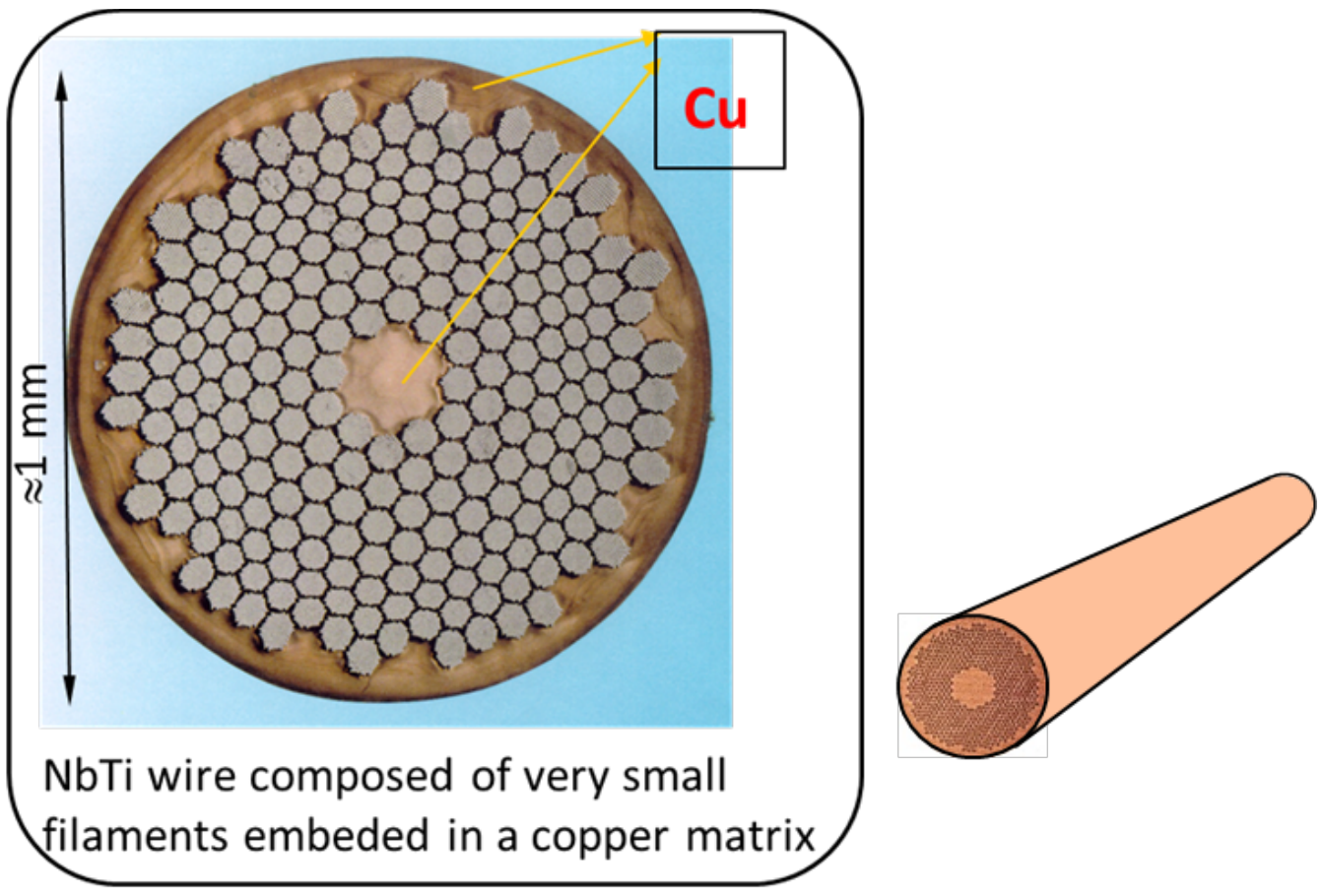

**Fig. 11:** Picture of a NbTi multifilamentary wire embedded in a copper matrix

### *3.3.3 Cabling and insulation process*

The cabling process involves assembling multiple superconducting strands into a compact, transposed structure known as a Rutherford cable (Fig. 13), using a specialized cabling machine. This process improves the performance and practicality of superconducting coils by reducing the strand length and number of turns, simplifying the winding, and lowering the coil inductance. The transposition of strands enables current redistribution during a quench, enhancing stability. The cable is flattened into a rectangular or trapezoidal shape to achieve a high packing factor (typically 88–92%) and maximize the efficiency of space usage in the coil. Key design parameters such as strand number, diameter, pitch angle, and compaction factor are carefully optimized during this stage. A cabling machine, such as the one shown in Fig. 12 from CERN, is used to twist and compact multiple superconducting strands (usually NbTi or $Nb_3Sn$) into a Rutherford cable, a flattened, transposed multi-strands cable. The cabling machine performs precise rolling, twisting, and shaping to ensure uniformity and mechanical integrity of the final cable.

The insulation process is the final step before coil winding, aiming to prevent electrical shorts between adjacent cable turns. Insulating materials must possess excellent dielectric strength, mechanical robustness under stress, and compactness for efficient winding. Additionally, they should allow helium or epoxy penetration (via porosity) and withstand radiation in demanding environments. In NbTi magnets, insulation typically consists of overlapped polyimide (e.g., Kapton™) tape, while $Nb_3Sn$ magnets often use fiberglass braiding. This insulation ensures the electrical and thermal integrity of the coil, especially during quench events or prolonged operation in cryogenic and high-radiation conditions. The insulation is applied with overlap (often 50%) to ensure uniform coverage and mechanical bonding during epoxy impregnation or operation in helium.

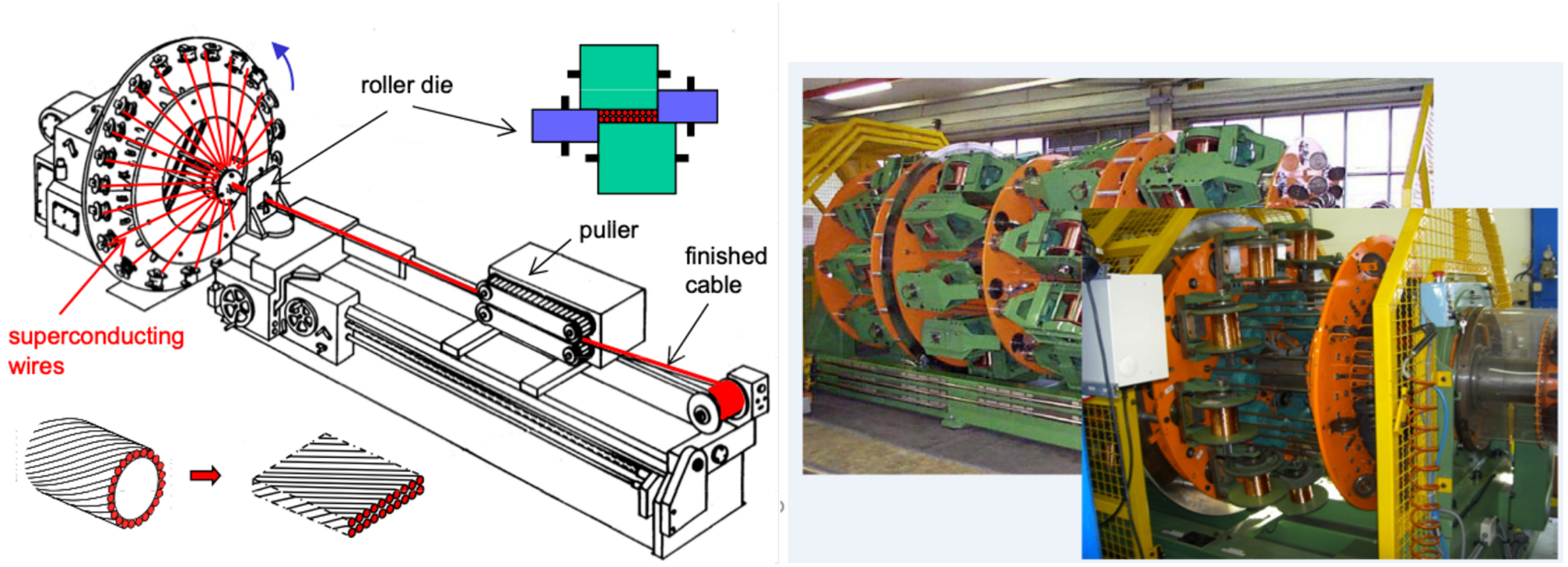

**Fig. 12:** Schematic and picture of a LTS strand cabling machine used at CERN [13].

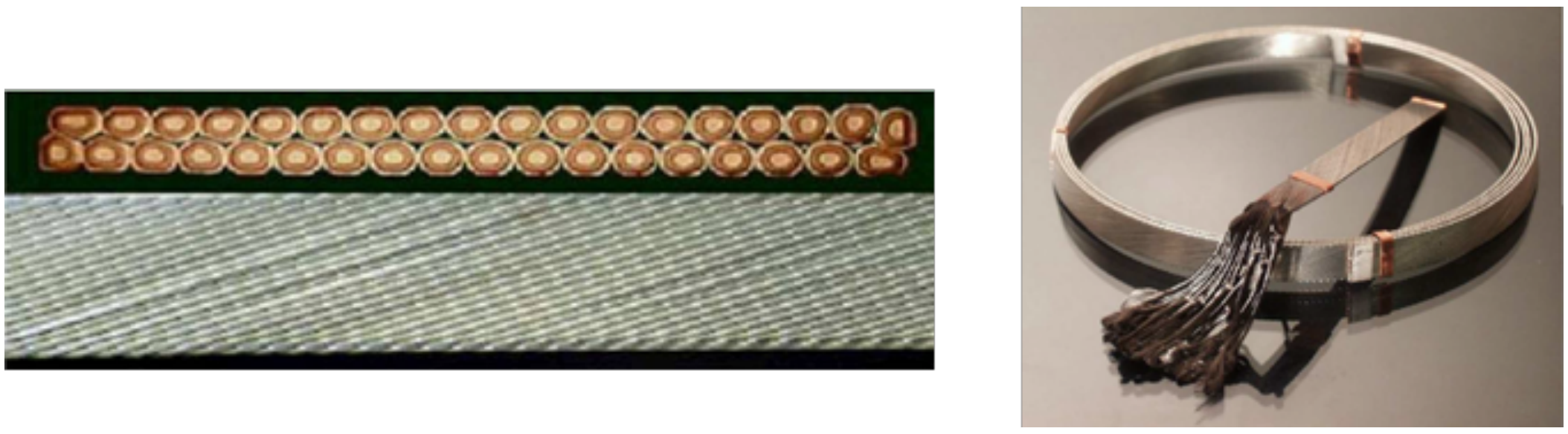

**Fig. 13:** Pictures of NbTi Rutherford cables

### *3.3.4 Short overview on optimized LTS strands and cables for accelerator magnets*

In accelerator magnets, low-temperature superconductor (LTS) wires or strands are designed with very specific characteristics to ensure high performance, reliability, and stability under intense electromagnetic conditions. These wires are typically multi-filamentary, containing thousands of extremely fine filaments (1,000 to 10,000) embedded in a copper matrix. Filament diameters range from 5 to 30 μm, which helps reduce magnetization effects and flux jumps, particularly important during injection and field ramping. The filaments are twisted (with a typical pitch of 25 cm) to minimize eddy currents and coupling losses. A carefully controlled Cu/non-Cu ratio (usually 1.5–2 ± 0.05) ensures both thermal stability and electrical performance. The wires must also be flexible for coil winding, have low AC losses, and be producible in long, cost-effective length.

Once these strands are twisted into Rutherford cables, further optimization is required for accelerator use. The cables must support high current levels (typically 10–20 kA) without significant degradation of the critical current density ($J_c$) compared to the individual strands. Uniform current distribution across the cable is essential, which demands precise dimensions, high filling factors, and controlled inter-strand resistance to manage current sharing during transients and quenches. The cables also retain the twisted configuration of the strands to reduce coupling losses during field ramping. This combination of fine filament design, optimal copper stabilization, and tightly controlled cabling makes LTS wires and cables well-suited for high-field accelerator magnets where performance and quench protection are critical. Table 1 summarizes the list of key parameters.

**Table 1:** Key parameters for particle LTS wire and cables [4]

| Multifilamentary wire (strand) | Cable |
|---|---|
| High and uniform $J_c$ at operating field | High-current cables (10 - 20 kA range) |
| Small filaments size to a) reduce magnetization and assure uniform field - mainly at injection, b) avoid flux jump | Minimum $J_c$ degradation with respect to virgin strands |
| Filaments twist to minimize coupling effects during ramping (eddy currents) | Uniform current density |
| Appropriate (Cu/non Cu) ratio - minimum amount of copper needed for stability and protection (typically 1.5-2 ± 0.05 for accelerator magnets) | High filling factor and ratio |
| Flexible, small bend radius (wound in coils) | Precise dimensions |
| Low AC losses | Twisted wires to minimize coupling effect during ramping |
| Low cost & long length | Controlled inter-strand resistance between crossing strands in the cable |

### 3.4 Practical HTS conductors

High-temperature superconductors (HTS), such as Bi2212, Bi2223, Y123 ($YBa_2Cu_3O_{7-x}$), (often symbolized by ReBCO ($REBa_2Cu_3O_{7-\delta}$)), are ceramic oxides with a perovskite-based layered crystal structure. These materials are electrical insulators in the normal (non-superconducting) state, yet they exhibit superconductivity in the copper oxide ($CuO_2$) planes that are stacked along the crystallographic *c*-axis (Fig. 14). The number of $CuO_2$ planes directly influences the superconducting transition temperature ($T_c$), which exceeds 77 K in many HTS materials, allowing operative conditions above 4.2 K and conduction cooled mechanism (Table 2). The structure is highly anisotropic: superconductivity and current transport are primarily confined within the *ab* planes, leading to a pronounced anisotropy in critical current density ($J_c$) and the upper critical field ($Bc_2$), with the irreversibility line (where flux pinning vanishes) lying well below $Bc_2$, especially along the *c*-axis. This anisotropy impacts vortex dynamics and pinning mechanisms, making the behavior of HTS materials very different from conventional (LTS) superconductors. For practical applications—particularly in high-field magnets—HTS materials are processed into long wires and tapes, often in the form of coated conductors or cables like CORC™, Roebel, or Twisted Stacked Tape Cable (TSTC) (Fig. 15). However, their ceramic nature and layered structure pose challenges for manufacturing: obtaining long, defect-free lengths with uniform properties is difficult, and weak links between grains or tape segments can significantly degrade performance. As a result, advanced fabrication techniques are essential to enable current sharing, reduce AC losses through transposition, and improve mechanical flexibility and bending capabilities. Multi-tape and cable architectures are key to overcome these limitations and unlock HTS potential in demanding applications such as accelerator or fusion magnets.

**Table 2:** Crystallographic parameters, number of adjacent $CuO_2$ planes and $T_c$ (H=0), [27].

| | Bi2212 | Bi2223 | Y123 |
|---|---|---|---|
| a [Å] | 5.415 | 5.413 | 3.8227 |
| b [Å] | 5.421 | 5.421 | 3.8872 |
| c [Å] | 30.880 | 37.010 | 11.680 |
| # of adjacent $CuO_2$ planes | 2 | 3 | 2 |
| $T_c$[K] | 91 | 110 | 92 |

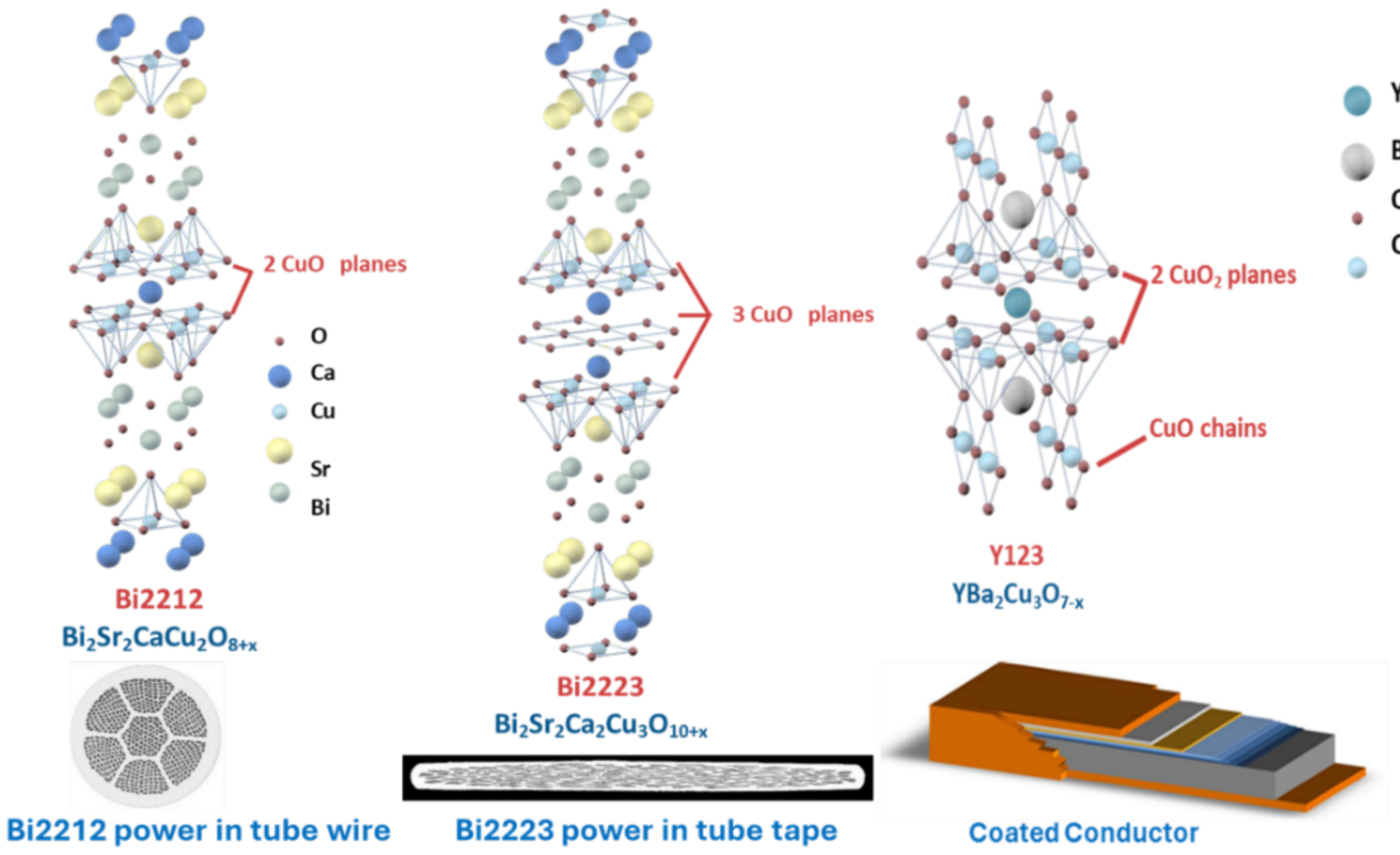

**Fig. 14:** Crystallographic layered and anisotropic structures of commonly used HTS material: Bi2212, Bi2223 and YBCO. The superconductivity appears in the $CuO_2$ planes [27].

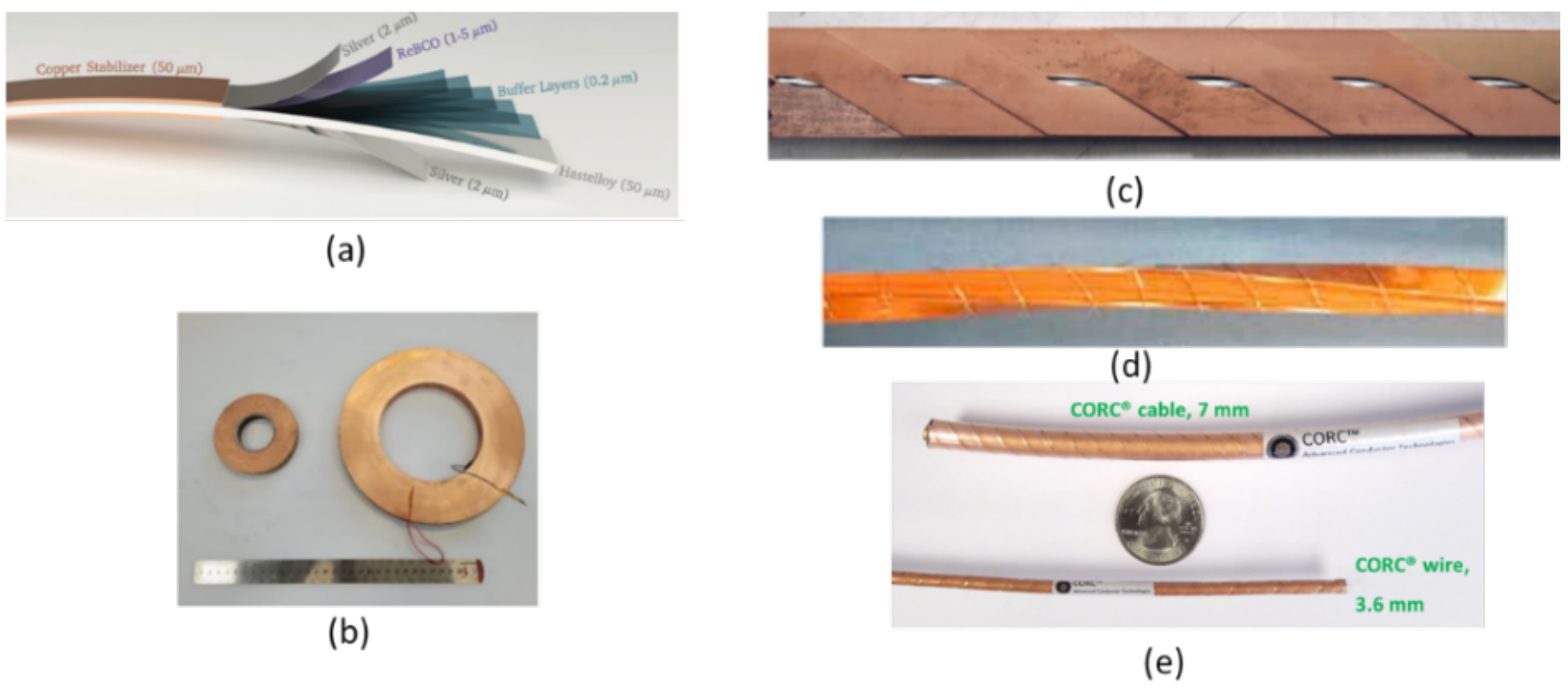

**Fig. 15:** (a) ReBCO coated conductor with an HTS film of 1-2 mm thickness deposited on one side of the substrat; (b) Non insulated HTS (bi) tapes used at PSI for a 15 T HTS solenoid [16]; (c) Roebel cable by the KIT assembled from Bruker ReBCO tapes- fiber glass rope inserted in the central channel [14]; (d) Twisted Stacked Tape Cable [29]. (e) CORC™ cables made of 3.6 mm and 7 mm wires [32].

The following table summarizes Low-Temperature Superconductors (LTS) vs High-Temperature Superconductors (HTS), focused on practical considerations like manufacturing, length, critical current, magnetic field performance, quench behavior, field quality, and cryogenic compatibility.

**Table 3:** Key aspects for practical LTS and HTS wires and cables

| **Aspect** | **LTS (e.g. NbTi, $Nb_3Sn$)** | **HTS (e.g. Bi2212, Bi2223, YBCO/ReBCO)** |
|---|---|---|
| Manufacturing complexity | Mature industrial processes; easier to fabricate (especially NbTi) | Complex ceramic processing; more fragile; Bi2223 and ReBCO need precise layering |
| Achievable conductor length | Long lengths (>km) routinely achievable | Still limited (100–500 m) for ReBCO |
| Mechanical flexibility | Good (especially NbTi); $Nb_3Sn$ is brittle after reaction | Brittle and strain-sensitive (especially Bi-based); ReBCO tapes are more robust |
| Critical magnetic field ($H_{c2}$) | ~15 T (NbTi), ~25–30 T ($Nb_3Sn$) | >100 T in ReBCO; HTS outperforms LTS above ~20 T |
| Critical current density at high field | NbTi: ~2500 A/mm² at 6 T, $Nb_3Sn$: ~3000 A/mm² at 12 T | Lower Jc at low field, but sustained Jc > 20 T |
| Quench protection | Fast normal zone propagation; well known quench detection and protection | Very slow quench propagation; quench detection and protection are challenging |
| Field quality (homogeneity) | Excellent; used in MRI and NMR with tight tolerance | Challenging due to tape geometry and anisotropy |
| Cooling temperature | Needs superfluid, liquid or supercritical He (1.9 to 4.5 K) | Operates at 20–77 K; enables cryocooler or $LN_2$-based systems |
| Cryocooler compatibility | Limited | Good; many HTS magnets operate with compact cryocoolers |
| AC losses | Moderate to low in twisted wires | High due to tape geometry; transposition needed (e.g., Roebel, TSTC) |
| Temperature margin | Small (few K) | High temperature margin due to higher operating temperature |

# 4 HTS magnets

## 4.1 Challenges and expectations from superconducting magnets

Normal conducting and superconducting magnets differ fundamentally in their physical principles, operational constraints, and design requirements. In normal conducting magnets, the magnetic field is shaped by the geometry of the coil and the magnetic properties of ferromagnetic materials (e.g. iron), with constraints due to resistive losses and thermal dissipation, which limit the achievable magnetic field and efficiency. In contrast, superconducting magnets rely entirely on the geometry of the current distribution, since superconductors exhibit zero electrical resistance below a critical temperature. This allows them to generate extremely high magnetic fields with no Joule heating. However, superconducting systems come with substantial challenges: their performance is inherently limited by the conductor properties (you cannot exceed the critical current density without a quench), and they are subject to mechanical constraints due to the large Lorentz forces at high fields. Quench protection becomes essential because a sudden transition to the normal state can cause rapid energy deposition and damage. Moreover, the magnetization of superconducting filaments can introduce hysteresis and field distortions, and at low fields, dynamic effects further degrade the field stability. The need for cryogenic cooling— using immersion bath, forced-flow or cryocoolers—adds significant complexity, and both production and operation are more expensive compared to room-temperature magnets. Handling cryogenic fluids, ensuring reliable electrical interconnections through cryo-lines, and maintaining stringent thermal conditions are all non-trivial tasks that make superconducting systems inherently more delicate and specialized.

Despite these limitations, the expectations for superconducting magnets are high and driven by their unmatched potential in high-field and high-current applications. From a materials perspective, superconductors are sought for their high engineering current density ($J_e$), and high irreversibility field ($B_{irr}$), which together enable reliable performance in intense magnetic environments. The engineering current density ($J_e$) is defined as the total transport current divided by the entire cross-sectional area of the conductor, including stabilizer, insulation— not just the superconducting material. The ideal superconducting material is not only high in $T_c$ but also exhibits low losses and stability across varied field and temperature conditions. On the fabrication side, flexible and scalable production methods are essential, allowing wires and tapes to be formed into long-length conductors with consistent properties. Mechanically, superconductors must be robust enough to handle electromagnetic stresses during operation. Cost-efficiency, though a challenge, is increasingly addressed by innovations that reduce material waste, improve cryogenic systems, and enable compact designs. Superconducting magnets are expected to be environmentally sustainable, thanks to their energy efficiency and minimal heat loss. Additionally, high-performance cables—like twisted stacked tape or CORC™ ones—allow for current sharing and mechanical flexibility. Ultimately, the goal is to achieve large-scale deployment of superconducting magnets with minimal operational overhead, enabled by efficient magnet protection, low ramping losses, and reliable performance. Applications are governed more by $J_c$, $J_e$, and $B_{irr}$ than by the critical temperature alone, making the engineering and material integration central to future advances in superconducting technology. The Venn diagram Fig. 16 illustrates the interconnected domains of technical difficulty, including screening currents, cabling and splicing, cost constraints, mechanical stresses, cryogenics, magnet protection, and electrical integrity. Each domain involves specific issues such as ramp losses, transposition requirements, field orientation, cryogenic management, and protection schemes, all of which must be addressed in parallel to achieve reliable and scalable magnet performance.

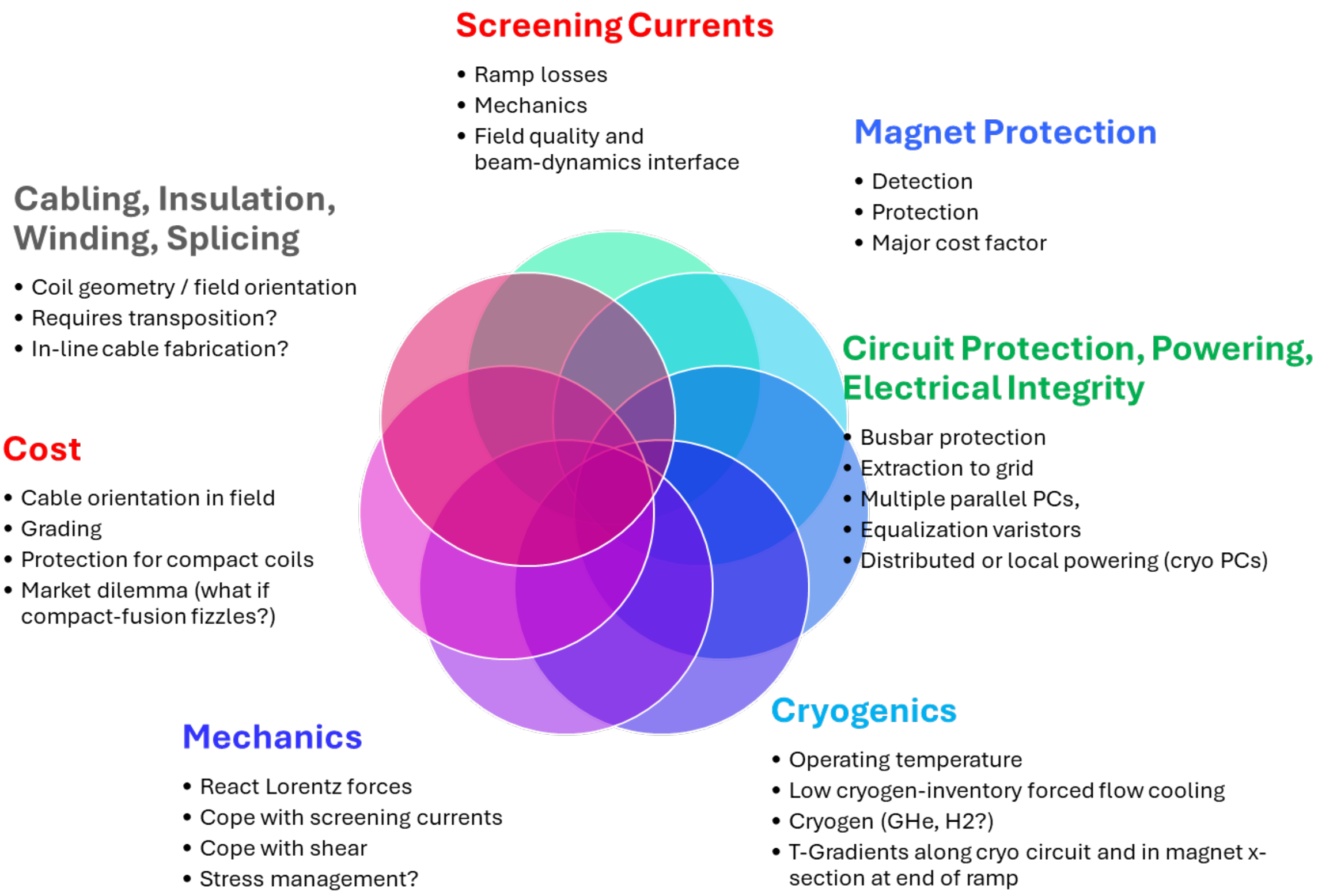

**Fig. 16:** Main technical challenges in the development of high-performance superconducting magnets [1].

### 4.2 Superconducting magnets: production process

The production process of superconducting magnets (Fig. 17) is quite different to what was presented in the previous course on normal conducting ones. Four phases are identified:

- Design: Involves multi-physics simulations (field quality, thermal, quench, CAD).
- Manufacturing: Focus on assembly control, material quality, impregnation, cryostat design.
- Qualification: Demanding tests for electrical, cryogenic, and magnetic performance.
- Installation & Maintenance: Procedure-focused, with high emphasis on traceability and safety.

Both superconducting magnets and normal-conducting magnets follow a similar structured life cycle—spanning design, manufacturing, qualification, and installation—with shared needs for rigorous specifications, CAD modeling, performance simulations, and quality control procedures. However, superconducting magnets differ fundamentally due to their reliance on cryogenic systems and the need for robust quench protection, making their design and operation significantly more complex. While both types undergo electromagnetic field quality testing, superconducting magnets require additional validation for cryogenic and electrical integrity, and their assembly involves delicate materials and cryostat integration. Maintenance and traceability are also more critical due to the higher investment, sensitivity to environmental conditions, and tighter operational margins. Thus, despite procedural similarities, superconducting magnets demand for more stringent monitoring and technical precision across all stages of their life cycle.

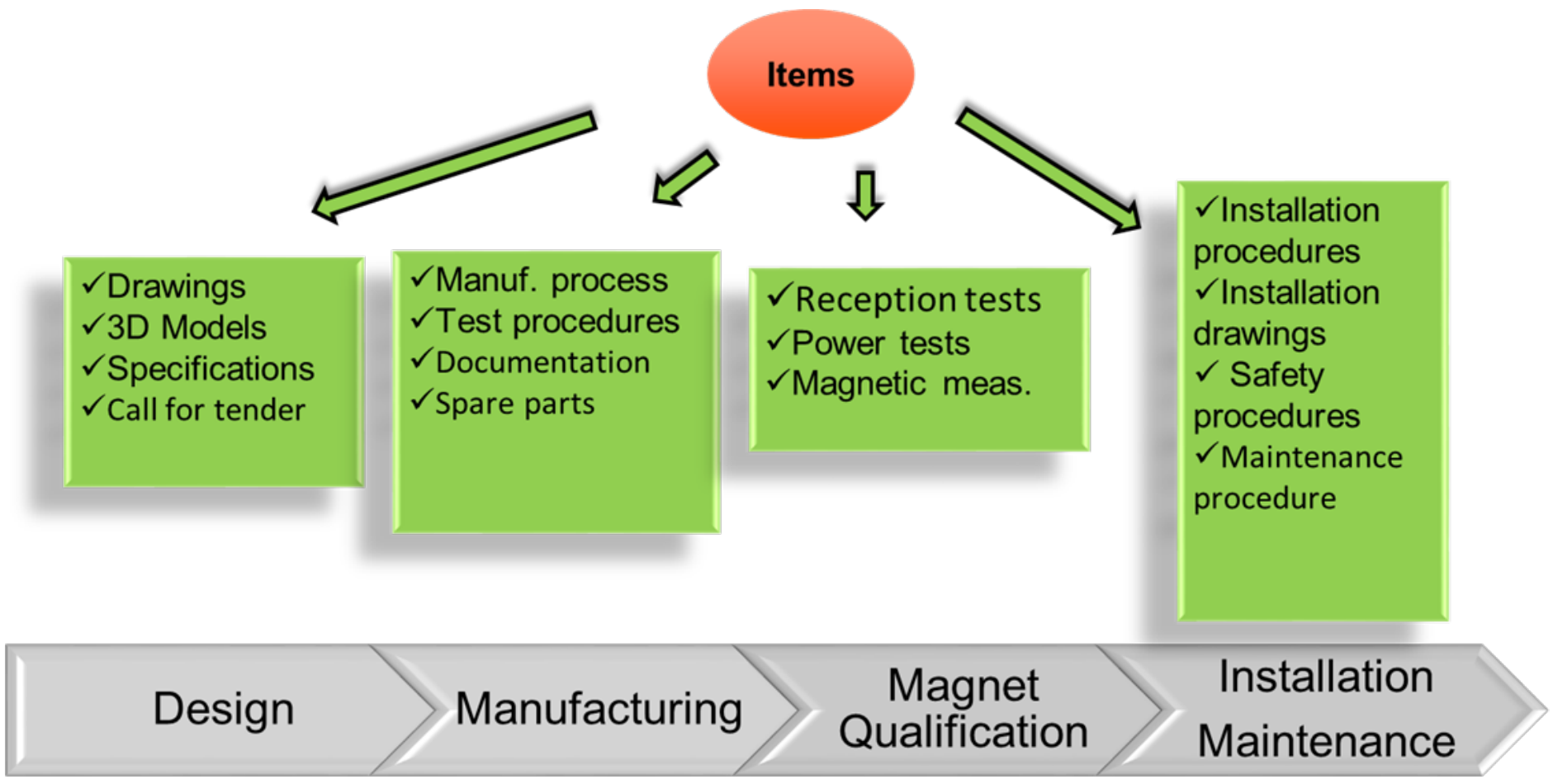

**Fig. 17:** Production and installation process of superconducting magnets

The following table summarizes the main similarities and differences with respect to normal conducting ones.

**Table 4:** Similarities and differences in Normal Conducting and Superconducting Magnet assembly.

| Items | Superconducting Magnets | Normal-Conducting Magnets |
|---|---|---|
| Design Simulations | Multiphysics: electromagnetic, thermal, hydraulic, mechanical, quench | Primarily electromagnetic and thermal simulations |
| Assembly Complexity | More complex (heat treatment, coil impregnation, cryostat integration) | Simpler assembly, fewer critical interfaces |
| Magnet Structure | Alloys or ceramic Materials (NbTi, $Nb_3Sn$, HTS) | Copper/Al conductors with robust mechanical support |
| Cooling | Requires cryogenic systems (LHe bath or cryocoolers) | Air- or water-cooled; simpler and cheaper |
| Quench Protection | Critical due to loss of superconductivity risk | Not needed |
| Testing | Electrical, magnetic, and cryogenic integrity tests | Electrical and magnetic tests; no cryogenic integrity |

# 5 Magnet manufacturing process

## 5.1 Overview

The manufacturing process of superconducting magnets is significantly more intricate than that of normal-conducting magnets due to the sensitivity of superconducting materials, the need for cryogenic operation, and the importance of quench protection. Each phase—from cable fabrication to final cryostating—requires extreme precision, rigorous quality control, and multidisciplinary integration. While both magnet types require accurate mechanical assembly and field quality, superconducting magnets involve additional thermal, electrical, and mechanical constraints.

*Multi-Wire Cable Fabrication & Insulation*

- Fabricate multi-filament superconducting wires (e.g., NbTi or $Nb_3Sn$).
- Assemble wires into Rutherford-type cables.
- Wrap with high-performance insulation (glass fiber, ceramic).
- Ensure electrical isolation and mechanical integrity.

*Coil Winding & Transition Manufacture*

- Wind the insulated cable onto mandrels with precision.
- Maintain proper tension and geometrical accuracy.
- Shape end turns and transitions carefully to avoid mechanical strain.

*Instrumentation Integration*

- Install voltage taps for quench detection.
- Embed temperature sensors (RTDs or thermocouples).
- Optionally include strain gauges for mechanical diagnostics.

*Coil Reaction (Wind & React for $Nb_3Sn$)*

- Heat-treat $Nb_3Sn$ coils at ~650–700°C in a controlled environment.
- Form superconducting $Nb_3Sn$ phase post-winding.
- Use heat-resistant insulation that withstands reaction temperatures.

*Vacuum Impregnation*

- Place coils in vacuum chamber and impregnate with epoxy or wax.
- Fill all voids, bond windings, and ensure mechanical rigidity.
- Cure resin thermally in a controlled cycle for optimal performance.

*Collaring Process (Stainless Steel Collars)*

- Compress coils under 50–100 MPa using a collaring press.
- Lock the stainless-steel collars with insertion keys.
- Pre-stress compensates for Lorentz forces during magnet operation.

*Yoking (Mechanical & Magnetic Integration)*

- Insert collared coils into a precision-machined iron yoke.
- Yoke enhances magnetic field return and adds structural support.
- Weld the yoke using high-force presses (up to 19000 tons).

*Cryostating (Cold Mass Insertion)*

- Insert magnet into vacuum-insulated cryostat.
- Apply multilayer insulation (MLI) blankets and thermal shields.
- Install liquid helium cooling circuits or cryocoolers.
- Ensure long-term thermal stability at cryogenic temperatures (~4.2 K, 20 K….).

## 5.2 Lorentz force containment and stress management

One of the most significant engineering challenges in superconducting magnet design—particularly in high-field applications like those in the LHC—is managing the enormous Lorentz forces generated by the interaction of high currents with strong magnetic fields. These forces act both radially outward and vertically inward, tending to deform and displace the superconducting coils. Specifically, the coils experience compressive forces in the vertical direction (towards the mid-plane, $F_y<0$) and expansive forces radially outward ($F_r>0$), which can reach several hundred tons per meter (e.g., $F\approx$ 340 tons/m in the longitudinal direction of the LHC dipoles). If not properly contained, these stresses can cause microscopic or macroscopic conductor motion, leading to mechanical instabilities, frictional heating, and ultimately quench. Coil deformation can also induce unwanted field errors. Therefore, effective mechanical containment is essential to maintain the structural integrity and stability of the magnet during energization and operation.

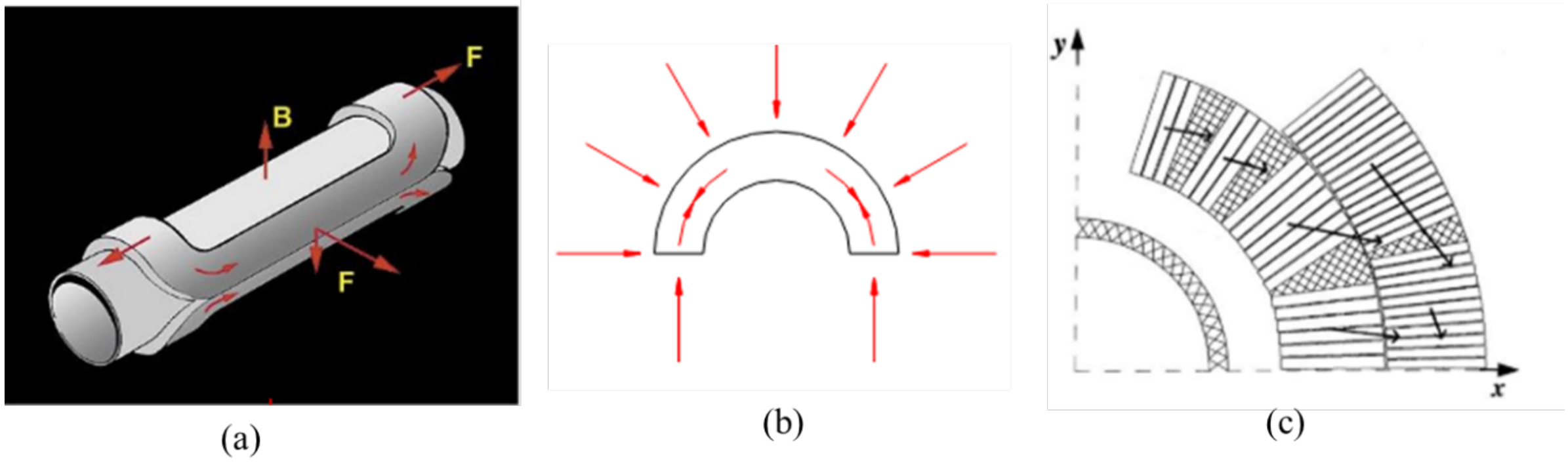

**Fig. 18:** Illustration of the various components of the Lorentz forces induced by high operating currents and produced magnetic fields [24].

To contain these forces several options are chosen already during the conception step:

a) Reinforcement with Mechanical Structures is a fundamental approach to counteracting the enormous Lorentz forces generated in high-field superconducting magnets. These forces, which tend to compress the coils toward the mid-plane and push them outward radially, can reach several hundred tons per meter, as seen in LHC dipoles. To contain such stress, the classical solution is to reinforce the structures with components like collars, yokes, shells, and outer cylinders that serve as mechanical support elements (Fig. 19 in the case of the LHC dipoles). These structures are carefully designed to distribute the forces uniformly across the coil pack and resist deformation. Thin collars are placed around the coil to enclose it tightly within a fixed cavity. The structure is compressed and secured using pins or keys. At room temperature (300 K), a prestress 2–3 times higher than the required cold value is applied to compensate for stress relaxation during cooldown. In high-field magnets, this prestress can become excessive; for example, in LHC dipoles, 70 MPa is applied at 300 K to achieve 40 MPa at cryogenic temperatures. The mechanical design of collars and yokes must account for all phases of magnet life—cooldown, energization, and operation—requiring detailed finite element simulations to model stress and strain under dynamic conditions.
b) Material Choices play a crucial role in withstanding mechanical loads without compromising structural integrity. The selected materials must not only be mechanically strong but also compatible with the extreme cryogenic environment. Reinforced composites, Inconel, titanium alloys, and G10 are commonly used due to their excellent strength-to-weight ratios, non-magnetic properties, and resistance to thermal contraction mismatches. Compatibility in thermal expansion between the support structure and the superconducting coil is essential to avoid the formation of mechanical gaps during cooldown, which could otherwise lead to coil motion, frictional heating, and eventual quench. The use of durable, cryo-compatible materials helps maintain tight dimensional tolerances, even under repeated thermal cycling and intense electromagnetic loads.

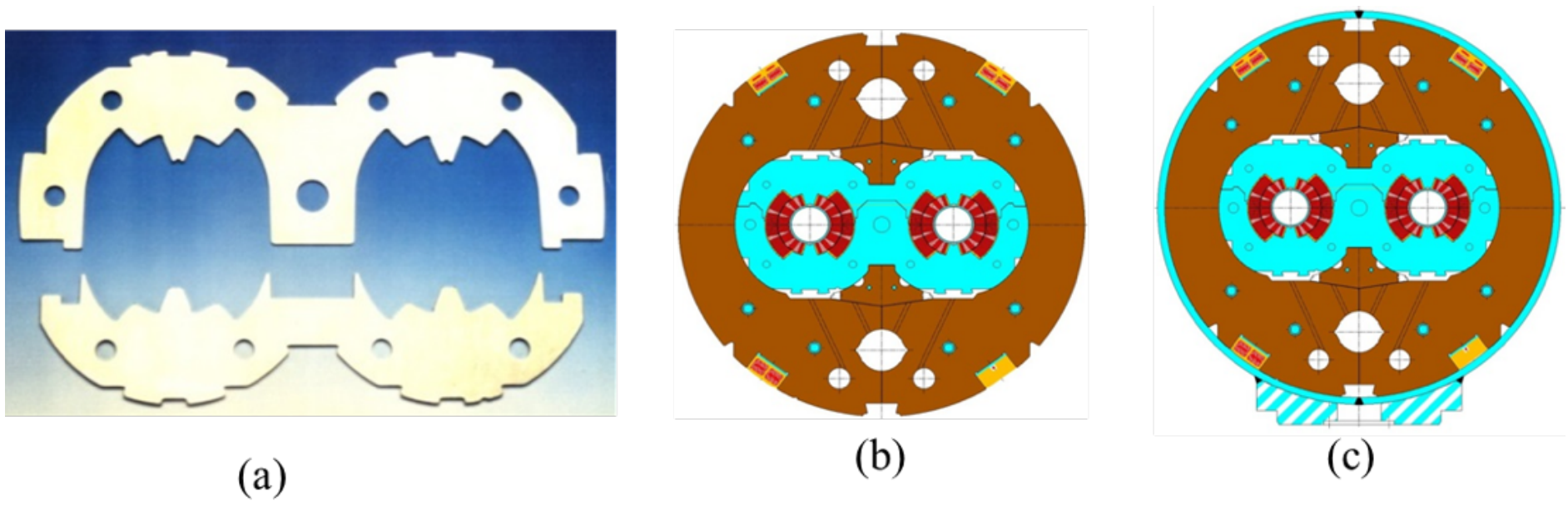

**Fig. 19:** Various components to reduce the mechanical instability caused by the Lorentz forces in the case of the twin aperture 8.3 T LHC dipoles: (a) stainless steel collars; (b) ferromagnetic yoke providing an increase of 15 % of the magnetic field and of the rigidity of the coil support structure, limiting radial displacement; (c) Shell welding with two half shells welded around the coil to provide additional rigidity [26].

c) Pre-stressing techniques are, aforementioned for the collars, among the most effective methods to ensure coil stability. By applying a well-calculated compressive preload during assembly, components such as the coils, collars, and shells are pre-compressed to counteract the expansive Lorentz forces during operation. This prevents the opening of gaps or the onset of tensile stresses that could cause motion or degradation. In LHC magnets, stainless-steel collars apply high-precision compressive forces to maintain coil alignment and prevent displacement. More advanced techniques, such as the key and bladder method, use inflatable bladders and mechanical keys to apply adjustable and uniform pre-stress. These methods ensure that the magnet remains mechanically locked even at full excitation. In superconducting magnets using $Nb_3Sn$ conductors, stress management is a critical aspect of design due to the material's brittle nature and strain sensitivity (see Chapter 3.2). Unlike NbTi, $Nb_3Sn$ material cannot tolerate significant mechanical deformation once reacted, as exceeding its strain limits can drastically degrade its current-carrying capability (critical current density, Jc). During energization, the Lorentz forces acting on the coils generate large internal stresses, which, if not properly contained and distributed, can cause cracking or irreversible performance loss in the $Nb_3Sn$ filaments. To mitigate this, stress-managed coil geometries are employed, where electromagnetic forces are intercepted and redirected away from the conductor blocks. This is achieved using a system of formers, rods, keys, and stainless-steel pads, which distribute the mechanical loads throughout the magnet's support structure rather than concentrating them within the fragile coils. The picture Fig. 20 describes the cross-section design for a stress managed Nb3Sn dipole demonstrator design and fabricated at PSI in the framework of the CHART Mag Dev program. As we move towards even higher-field magnets—particularly those employing High Temperature Superconductors (HTS) like ReBCO or Bi-2212—the importance of stress management becomes even more pronounced. These materials can operate at fields exceeding 20 T, but the Lorentz forces scale with both current and field strength, leading to extremely high mechanical loads. Moreover, HTS conductors, especially tapes like ReBCO, are anisotropic and delamination-prone, requiring precise mechanical support to avoid transverse tension or shear. In this context, the coil geometry must be designed from the outset to intercept and distribute stresses, often incorporating specialized structures such as bladders and keys, modular coil blocks, and layered stress management systems.

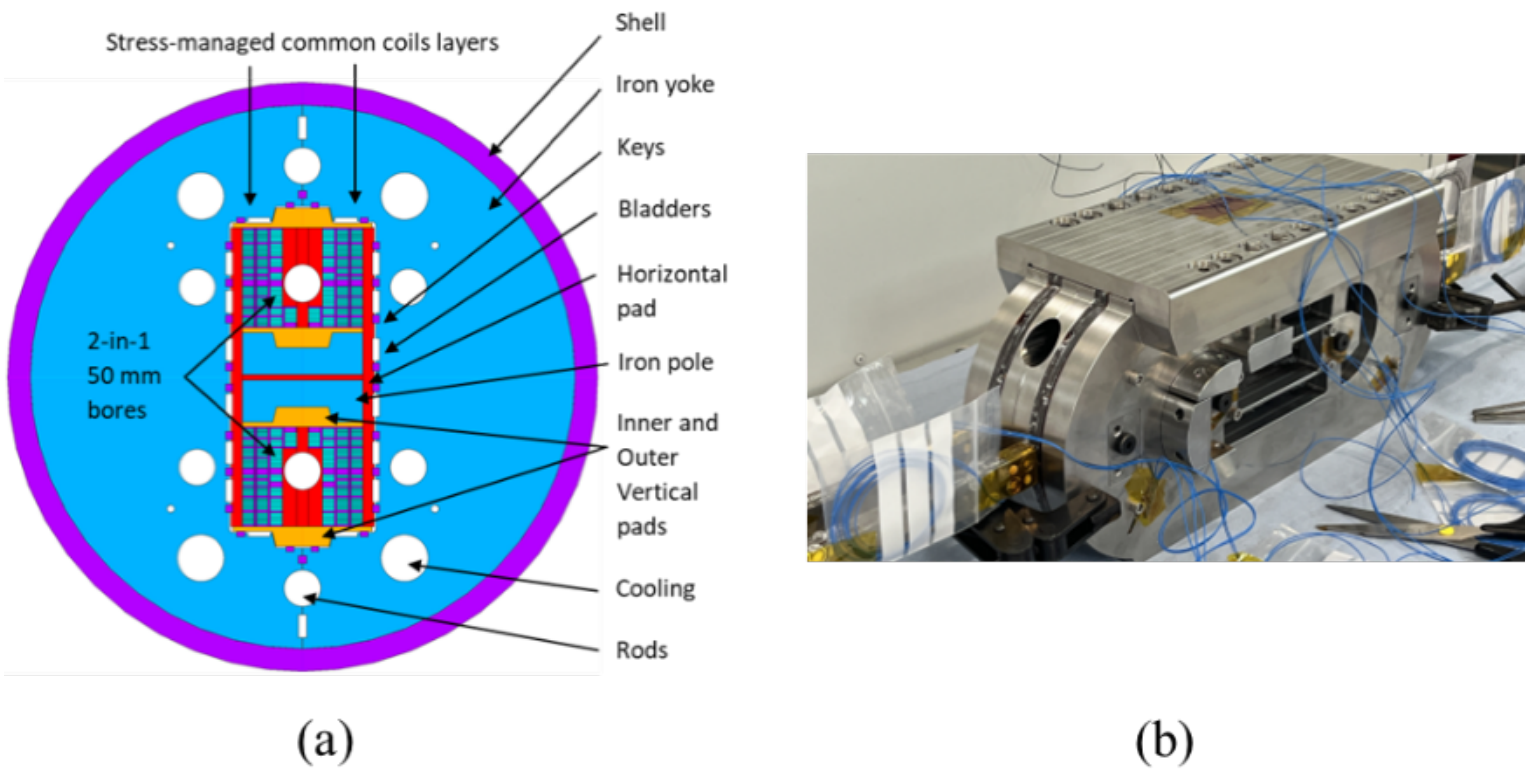

**Fig. 20:** Cross section and magnet prototype of a stress management Nb3Sn dipoles fabricated at PSI [3].

The most advanced and structurally robust concept currently explored for stress mitigation is the Canted Cosine Theta (CCT) geometry. In this design, two tilted solenoidal layers are wound such that each turn is individually supported by a mechanical structure with a mandrel and ribs and spars (Fig. 21).

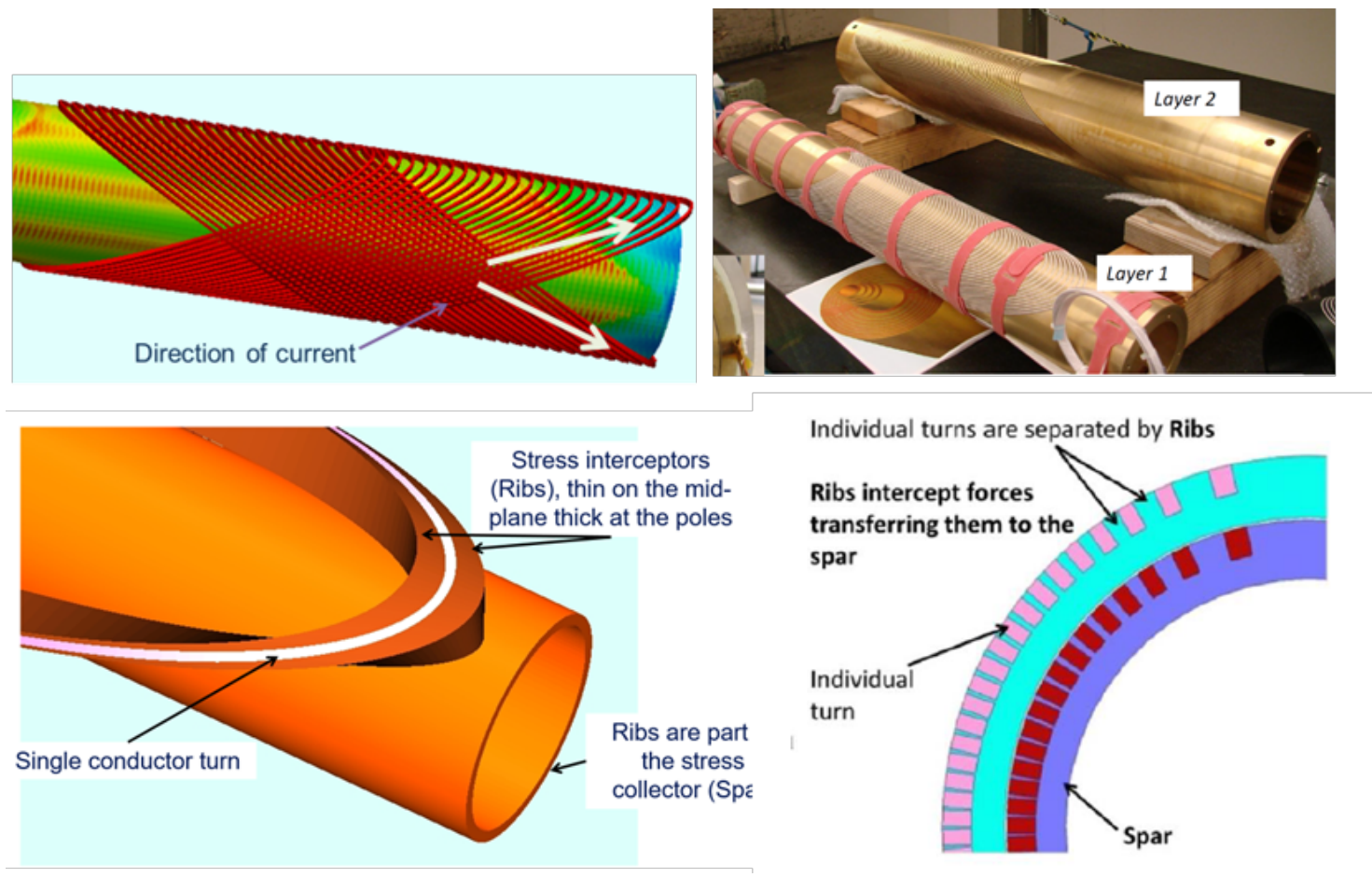

**Fig. 21:** Two layers of conductors made with Canted Cosine Theta (CCT) geometry. The effect of the Lorentz force of each turn is intercepted by the structure itself including ribs and spars [9].

These internal structural elements intercept the Lorentz forces locally at each turn and redirect them into the spar, preventing force accumulation and minimizing global mechanical strain. This approach eliminates the need for traditional collars, end parts, and spacers. The result is a mechanically stable configuration where stress is absorbed and managed turn-by-turn, making it ideal for very high-field applications, including future HTS-based magnets. CCT magnets represent an extreme case of integrated mechanical-electromagnetic design, where the geometry inherently supports the conductor and ensures robust performance under severe operational loads. Figure 22 illustrates a 1-m long, four layers dipole made of $Nb_3Sn$ cables wounded around a cylindrical Aluminum Bronze mandrel designed and fabricated at PSI [19]. The dipole was tested at CERN at 4.2 K and 1.9 K, reaching at the end of the training, 10.1 T in the bore at 1.9 K and 9.9 T or 100% of maximum field at 4.5 K [2].

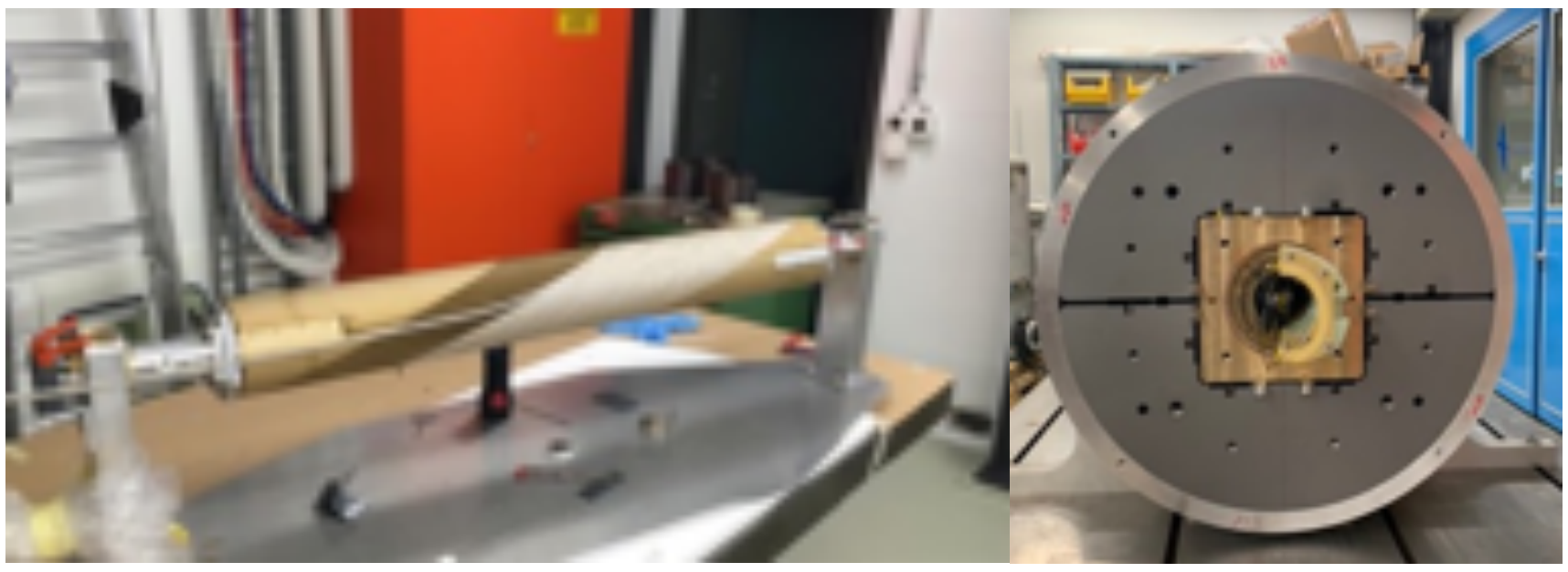

**Fig. 22:** Winding of one layer with Nb3Sn cable and assembled dipoles of a one 1 m long CCT dipole fabricated at PSI. Details of the manufacturing process in [19].

### 5.3 Magnet protection against quenches

A superconducting magnet that is not properly protected against quenches is at serious risk of catastrophic failure. During a quench, the sudden transition from the superconducting to the normal resistive state leads to rapid energy dissipation of the stored magnetic energy ($1/2\ LI^2$) as heat, which can cause severe overheating of the coil. As example, for the LHC superconducting dipoles, the stored energy at nominal field is 7.1 MJ which causes the melting of 13 kg of copper. This thermal surge can result in insulation degradation, conductor damage, and in extreme cases, even the meltdown of splices or the conductor itself. As the quench propagates, high voltages may develop across sections of the coil, potentially causing electrical arcing, short circuits, or breakdown of dielectric components. Additionally, the large and uneven thermal expansion associated with rapid heating introduces intense mechanical stresses, which can lead to structural failure or strain-induced cracking of brittle superconducting materials such as $Nb_3Sn$. Without adequate quench detection and protection systems, including fast discharge circuits and energy extraction mechanisms, a quench event can permanently damage the magnet, endanger nearby equipment, and compromise operational safety.

It is essential to detect the onset of a quench as early as possible. Early detection prevents the quench from spreading uncontrollably and allows protective systems to intervene before the energy stored in the magnet is released in a highly concentrated area. When a quench begins in a superconducting magnet, a resistive voltage rapidly develops across the affected region, marking the transition from the superconducting to the normal state. This voltage rise is a critical indicator used for quench detection. The quench protection system continuously monitors the voltage across the coil, and once it exceeds a predefined threshold—typically around 100 mV—the system initiates a multi-step protection sequence, the cut of the power supply and the triggering of the protection. To avoid false positives caused by noise or transient events, a validation time $\Delta t_{val}$ is implemented after the threshold is crossed to confirm that the signal corresponds to a true quench. Following validation, the system proceeds with the switching time $\Delta ts_{witch}$ to activate the protection mechanism—usually by inserting an external dump resistor—thereby increasing the total resistance in the circuit.

The magnet can be seen as a *L*/*R* circuit with a current decay following the exponential law $I(t) = I_0 e^{-t\frac{R(t)}{L}}$, where R(t) consists of both the energy extraction resistance $R_{EE}$ and the growing internal resistance from the quench zone $R_{quench(t)}$. The efficient protection consists in dissipating the stored magnetic energy safely over a reduced decay time $\Delta t_{dec}$ — the characteristic time over which the magnet current exponentially decreases once the protection is triggered. To speed up the decrease of the current inside the magnet the key objective is to increase the growth rate of the normal zone—the region of the coil that transitions from superconducting to resistive. Rapid expansion of this zone helps distributing the energy uniformly, preventing localized overheating and damage. Several strategies are adopted [17] – three methods are presented in the following:

- *Dump resistor activation*: Inserting an external dump resistor $R_{dump}$ (>> $R_{quench}$) into the circuit immediately after quench detection accelerates the current decay by increasing the total circuit resistance. Part of the energy is extracted, dissipated outside the magnet.
- *Quench heaters activation:* Heaters are resistive strips placed close to the superconducting coil. Quench heaters are made of stainless-steel strips (typically 25 µm thick) mounted on a polyimide sheet (typically 50 µm) and coated with copper cladding (~10 µm) to reduce voltage during operation. Once a quench is detected, the heaters are activated to intentionally raise the temperature of adjacent coil regions, forcing them into the resistive state. This method ensures symmetric and widespread growth of the normal zone and avoids hot spots. It's widely used in LHC dipoles and quadrupoles. While being an efficient system quench heaters have several drawbacks. Their response is delayed due to reliance on thermal conduction, and they often provide limited coverage, especially in complex or deeply embedded coil regions. Integration adds design complexity, and heater failure—due to electrical faults or insulation issues—can compromise magnet safety. Additionally, they may cause non-uniform quench propagation, require dedicated energy and trigger systems, and can degrade over time with thermal cycling, affecting long-term reliability. These limitations have led to the development of complementary methods such as CLIQ for faster and more uniform quench initiation.
- *CLIQ System (Coupling-Loss Induced Quench):* CLIC is an innovative and efficient quench initiation technique [23]. It works by discharging a capacitor into the magnet coils, generating fast-changing currents that induce coupling losses in the superconducting strands. These eddy-current-like losses quickly heat up the conductor volume, promoting a rapid and homogeneous transition to the normal state. Unlike heaters, CLIQ does not rely on thermal contact but instead uses electromagnetic coupling, making it faster and effective even in regions where heaters are hard to install. One major challenge is however the generation of high voltages, especially in long or large magnets, which can pose a risk to insulation integrity and require more robust and complex electrical design. The rapid oscillating current used to induce coupling losses can lead to voltage spikes across coil segments, making CLIQ less suitable for systems with tight voltage constraints or very long magnets unless additional voltage management strategies are implemented.

### 5.4 Infrastructure for magnet assembly

To conclude this chapter dedicated to magnet manufacturing, it is important to stress that the assembly of superconducting magnets is a highly complex and a multidisciplinary industrial process, requiring specialized infrastructure and precise coordination of numerous stages. The image in Fig. 23 shows the building 180 at CERN, a large integrated facility where each segment of the magnet manufacturing—from winding to final testing—is supported by dedicated, high-precision equipment.

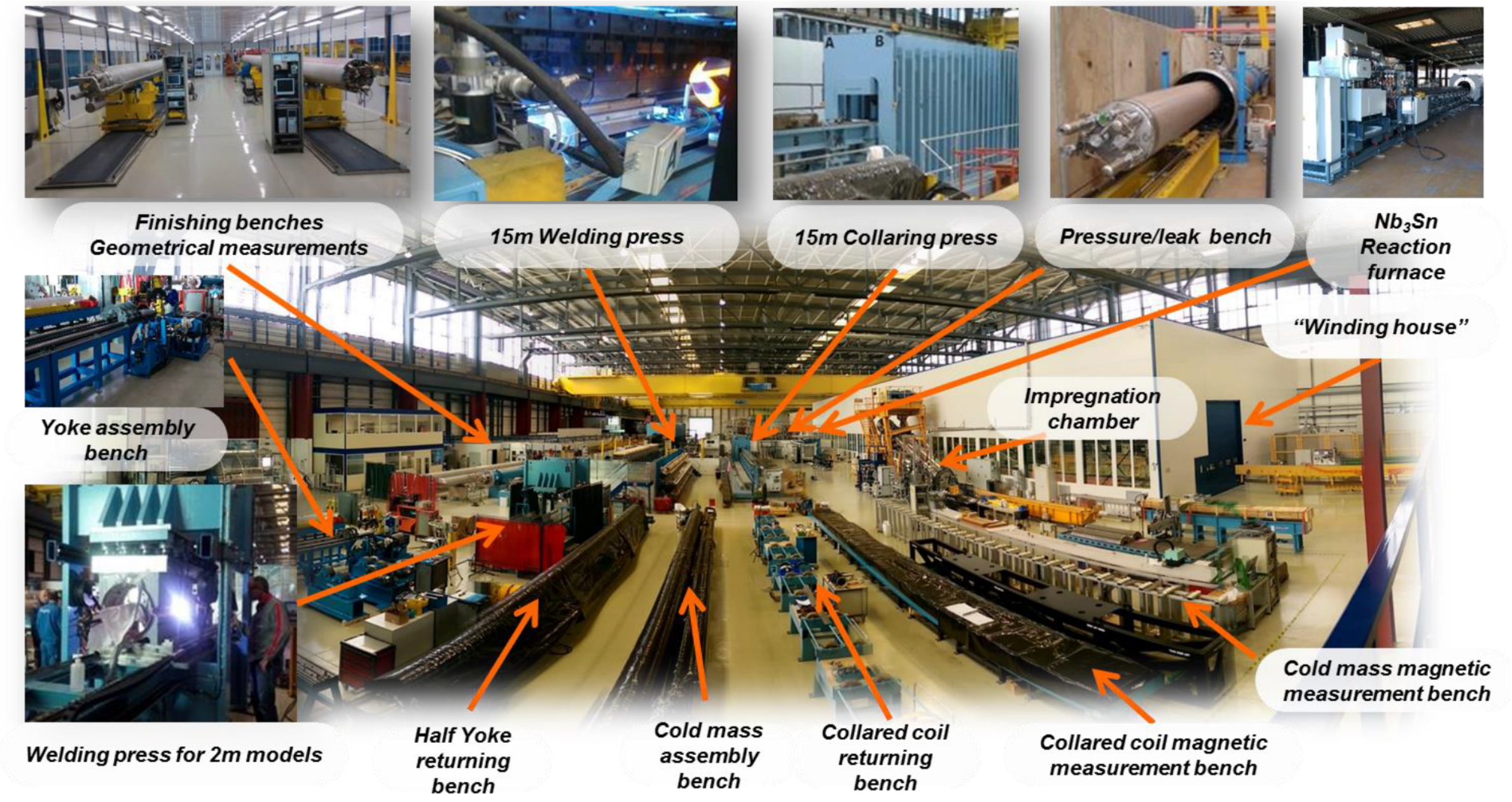

**Fig. 23:** Infrastructure at CERN for superconducting magnet assembly [26].

The process begins in the "winding house", where superconducting cables are wound into coils. For $Nb_3Sn$ magnets, these coils must be subjected to a high-temperature reaction process ($Nb_3Sn$ reaction furnace) to form the superconducting phase. This is followed by impregnation, where coils are vacuum-impregnated with epoxy resin to stabilize them mechanically. Once cured, coils are measured and aligned on collaring presses—15-meter-long machines that apply high mechanical pre-stress with micron precision to lock the coils into place. The welding presses, both for full-length (15 m) and short models (2 m), are used to assemble and seal key structural elements. After collaring, components move through a sequence of returning benches for repositioning and assembly, such as the collared coil returning bench and cold mass assembly bench. Magnets then undergo yoke assembly, structural welding, and multiple rounds of geometrical and magnetic measurements, using dedicated benches operating at room temperature. Pressure testing ensures vacuum and helium circuit integrity (pressure/leak bench), while cold mass magnetic benches verify field quality and alignment.

### 5.5 Four selected examples of superconducting magnets

Four examples of superconducting magnets using various materials and for different purposes are presented:

#### *5.5.1 The LHC cryo-dipoles*

The LHC cryogenic dipoles are highly sophisticated superconducting magnets designed to guide and bend proton beams within the Large Hadron Collider. Each dipole features a double-aperture configuration, allowing two counter-rotating beams to circulate simultaneously within a shared magnetic structure. The main specifications of an LHC dipole and its cross-section are illustrated in Fig. 24.

| | Main Parameters |
|---|---|
| 56 mm | Bore diameter |
| ~16.5 m | Total length |
| 1.9 K | Operating temp. |
| 8.3 T | Magnetic field |
| 11850 A | Nominal current |
| ~ 30 t | Mass |

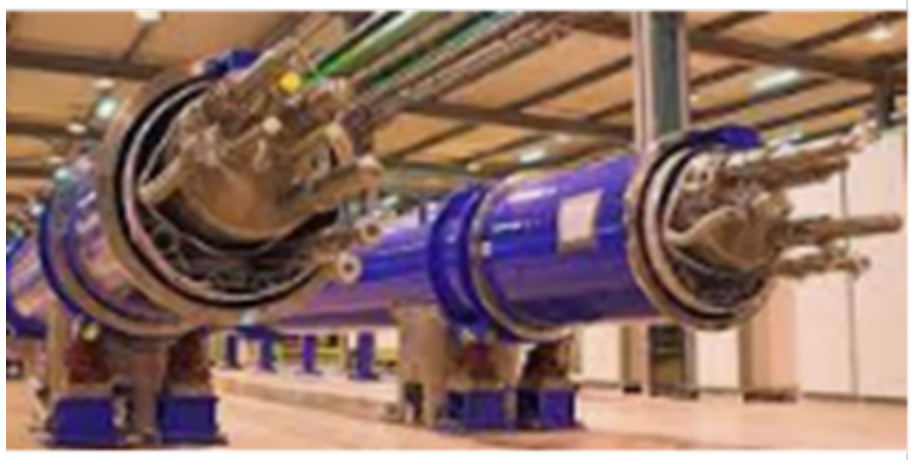

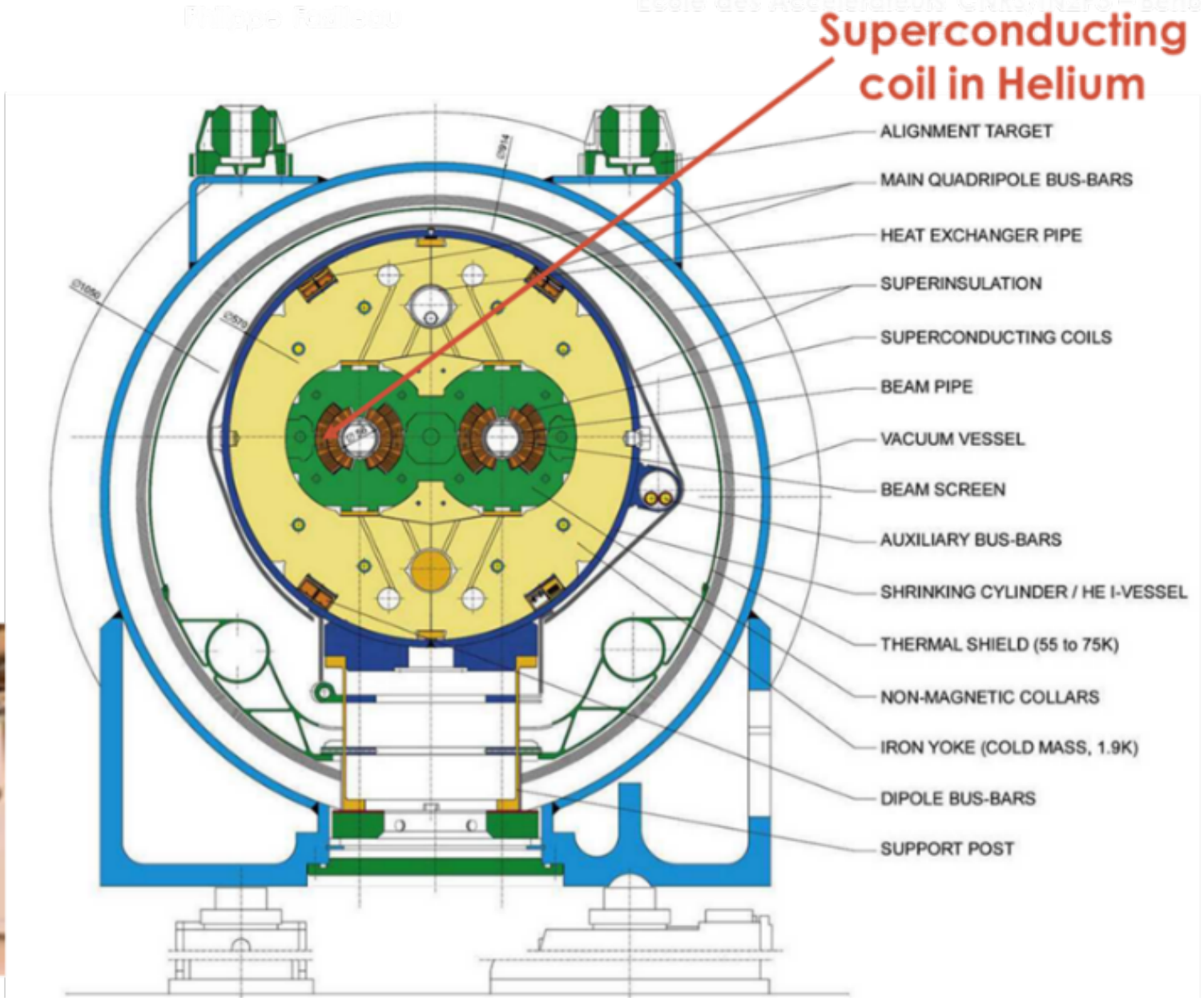

**Fig. 24:** Operating parameters and cross-section of the LHC superconducting dipoles [24].

It includes a bore diameter of 56 mm, a total magnet length of approximately 16.5 meters, an operating temperature of 1.9 K (superfluid helium), and a nominal magnetic field of 8.3 T generated by a current of 11850 A. Each magnet has a total cold mass of around 30 tons. The magnetic field is generated by NbTi superconducting coils wound in a cosine-theta geometry, a classical configuration that ensures excellent field uniformity and efficient use of superconducting material. This geometry produces a transverse dipole field by varying the current density as $J=J_0\cos\theta$, where $\theta$ is the angular position in the coil. While this configuration is optimal for field quality and conductor usage, it causes significant electromagnetic force buildup at the mid-plane, where both the mechanical stresses and the magnetic field reach their maximum. Consequently, the coil ends must be carefully designed to handle the large Lorentz forces and minimize stress concentrations. The coil is stabilized mechanically through a series of non-magnetic collars and is supported within a massive iron yoke that provides structural rigidity and field return. The structure is enclosed in a thermal shield and a shrinking cylinder and immersed in superfluid helium for optimal cooling at 1.9 K. The entire cold mass is enclosed in a vacuum vessel to ensure thermal insulation. The twin-aperture design is housed within a common collaring and yoke assembly, ensuring mechanical symmetry and efficient space usage in the LHC tunnel (Fig. 24). The dipole is protected using quench heaters along the coil length.

### 5.5.2 *The 5T superconducting superbends for the upgrade of the Swiss Light Source*

The two 5 T superbends developed for the SLS2.0 upgrade at the Paul Scherrer Institute are conduction-cooled superconducting dipoles designed to enhance hard X-ray production on two selected beamlines [8]. These magnets operate between 3 T and 5 T at 4.2 K, with a peak field centered on the beam axis, enabling higher photon energies and brightness. Each dipole features a compact, closed magnetic structure composed of two pairs of Nb-Ti superconducting coils—one inner and one outer—using a racetrack or solenoidal layout to achieve the desired field profile. The field distribution is highly optimized, with simulations confirming sharp field peaks and high uniformity in the beam region. For compactness, the dipoles are conduction-cooled using two cryocoolers (RDK-415D), eliminating the need for liquid helium baths. The superconducting coils are mounted within an aluminum pre-compression ring for structural support and embedded in a robust Armco closed iron yoke to shape the magnetic field.

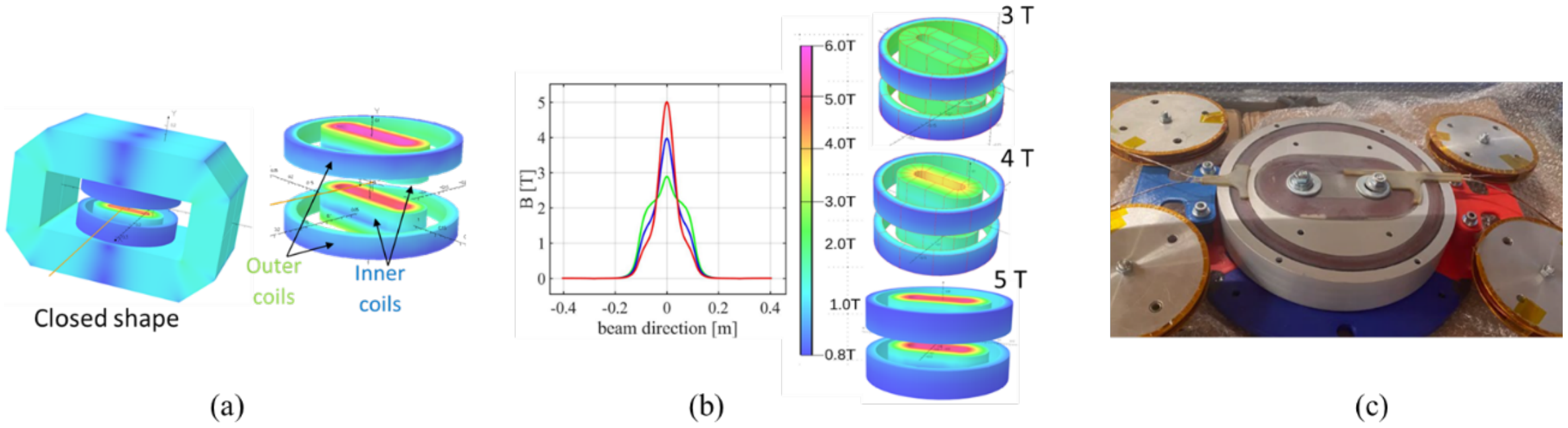

**Fig. 25:** (a) Field distribution of the magnetic field in the inner and outer NbTi coils and inside the closed yoke; (b) Longitudinal intensity of the magnetic field along the electron beam. (c) Picture of the first NbTi coils.

Additional components include Copper and HTS current leads, thermal connections, and suspension straps to stabilize the cold mass (Fig. 26). The dipoles are protected with external dump resistors [8].

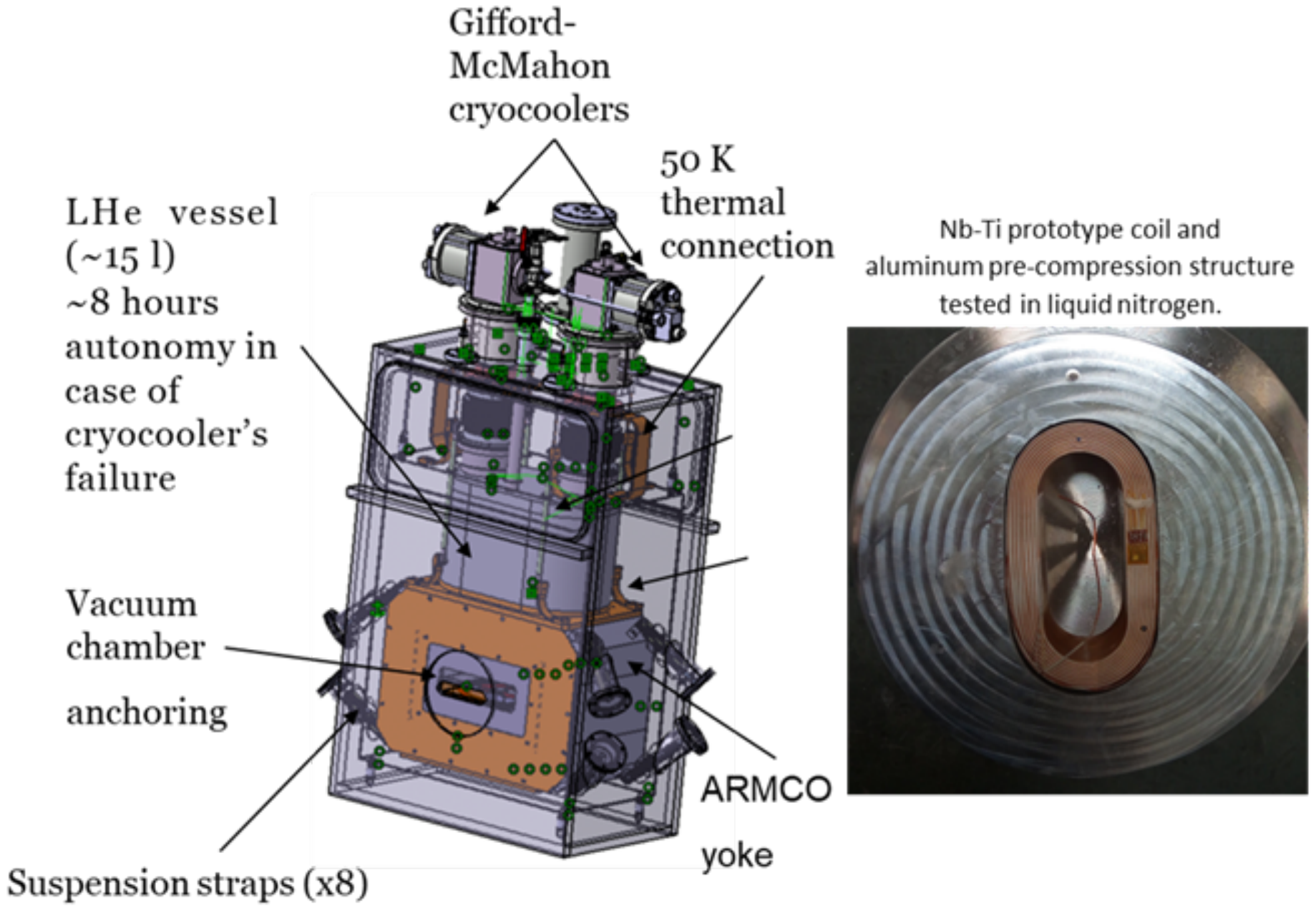

**Fig. 26:** Assembly design of the SLS2.0 superconducting superbend with a zoom on a NbTi superconducting coil [8].

### *5.5.3 15 T superconducting solenoid for the Positron Production experiment at PSI*

The high-field HTS solenoid developed at PSI for the FCC-ee positron injector demonstrator is an in-house magnet designed to reach intense magnetic fields in a compact geometry. As shown in Fig. 27, it consists of four stacked ReBCO (Rare-earth Barium Copper Oxide) pancake coils wound without insulation (NI), a strategy that allows lateral current sharing during a quench, enhancing the coil's stability and protection. Each coil has 170 turns and uses two ReBCO tapes, for a total conductor length of 2 × 49 meters. The solenoid features a 100 mm outer diameter and a 50 mm aperture, with a target peak field of 15 T in the bore, though test campaigns have shown it can achieve up to 18.2 T while maintaining temperatures below 12 K for over 3 hours. The magnet assembly is carefully designed with robust copper and stainless-steel mechanical support structures that provide pre-stress and stability. Thermally conductive interfaces ensure good heat extraction, and multiple voltage taps and sensors are installed for detailed monitoring. The absence of turn-to-turn insulation in the coils enhances the coil's thermal resilience and allows for fast redistribution of current during a disturbance. This architecture, though sensitive to mechanical tolerances, is particularly beneficial in high-energy applications requiring steady state strong fields with enhanced operational safety and reduced quench risks. In the test configuration, the HTS solenoid was housed inside a cryostat equipped with conduction cooling (no liquid helium bath), using two cryocoolers to maintain low operating temperatures (Fig. 27). The system integrates advanced diagnostics such as Hall sensors, fiber optics, and voltage taps for precise characterization of electromagnetic behavior. The NI HTS architecture, combined with high thermal conductivity materials and reliable mechanical integration, demonstrates the capability to engineer next-generation magnets for compact, energy-efficient accelerator technologies [16].

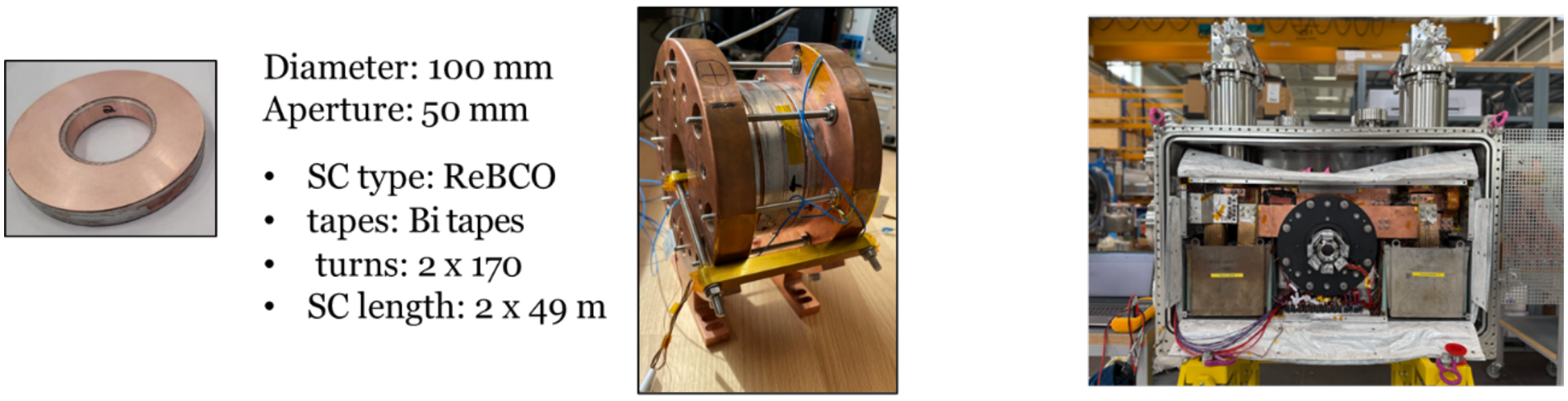

(a) (b) (c)

**Fig. 27:** 15 T conduction cooled superconducting solenoid (c) made of a stack of four Non Insulated HTS bi-tapes (a, b) used for the position production project at PSI [16].

### *5.5.4 Efficient Superferric Dipole demonstrator with $MgB_2$ coils*

Superferric dipole magnets represent a promising class of superconducting magnets characterized by a strong reliance on the geometry and magnetic properties of the iron yoke, which plays a dominant role in shaping the magnetic field. Unlike traditional high-field superconducting magnets, superferric designs use the iron flux return as the primary element to guide and define the field profile, while the superconducting coils—typically operating at moderate fields (~2 T)—serve mainly to excite the magnet. This configuration enables very efficient use of superconducting material, reducing both cost and energy consumption compared to conventional resistive magnets. A key advantage of superferric magnets is their simplicity and compactness: the close coupling between the coil and the iron yoke allows designers to minimize the cross-section and weight of the magnet, making them ideal for applications where space and infrastructure are limited. The example shown, developed at CERN, is a superferric dipole demonstrator using $MgB_2$ superconducting cable, operating at 4.2 K, with 5 kA current and delivering a magnetic field of 1.95 T (Fig. 28). This system highlights the feasibility of using cost-effective superconductors in cryogen-free or conduction-cooled configurations. Superferric magnets are also valued for their reliability, mechanical flexibility, and reduced fabrication and operative costs, making them suitable candidates for next-generation accelerator applications where moderate fields are sufficient. As a practical alternative to resistive electromagnets, they offer a balance between performance and operational efficiency—especially in scenarios where field quality is less critical or can be optimized via the iron geometry.

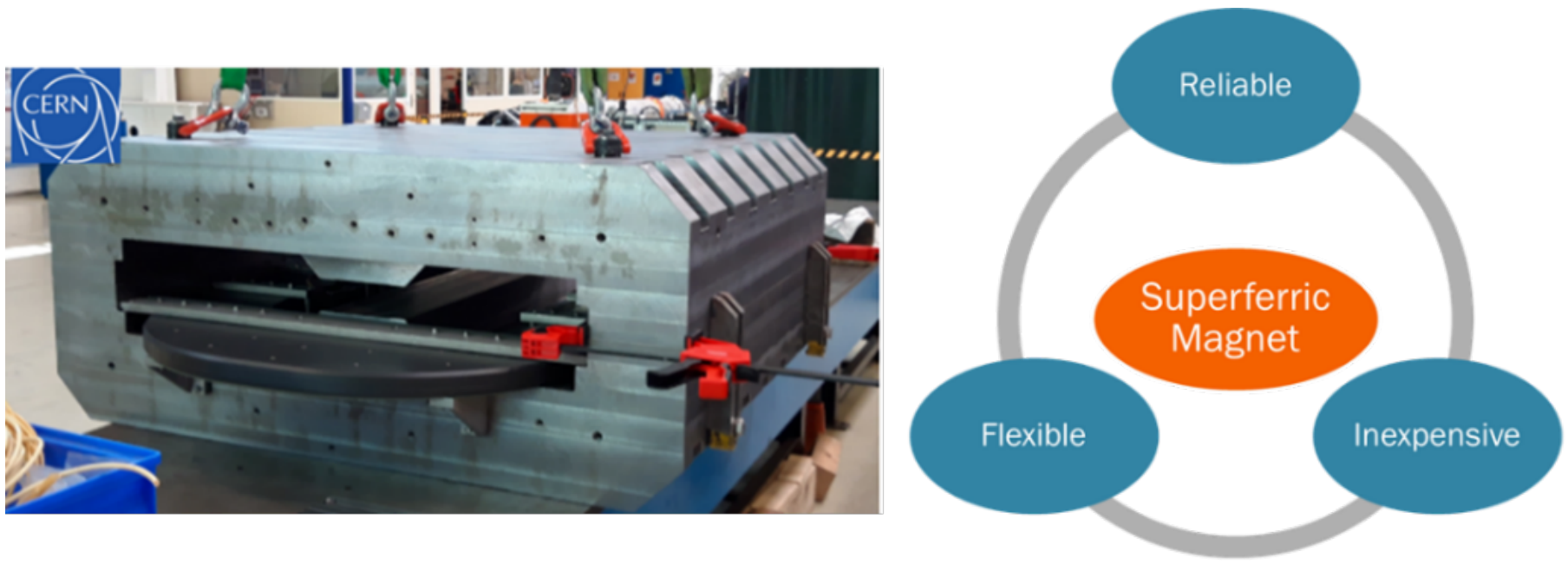

**Fig. 28:** Energy Efficient Superferric Dipole demonstrator with $MgB_2$ coils [12] and advantages of this type of magnet [28].

## 6 Superconducting magnet qualification at cryogenic temperature

### 6.1 Qualification at 1.9 K: example of the LHC cryo magnets

Qualification of superconducting magnets in operating conditions is a critical step in ensuring their performance, stability, and safety in accelerator environments. This process involves testing the magnets at cryogenic temperatures—typically at 4.5 K or lower, down to 1.9 K for high-field applications—to replicate actual operating conditions. The tests are designed to assess several key aspects: the electrical and cryogenic integrities, the current-carrying capability, the quench performance, and the effectiveness of protection systems and the magnetic field quality at steady state current or at various ramp rates. These tests include powering the magnet to nominal and above-nominal currents to verify stability, identifying any training behavior (progressive increase in quench current), performing field integral and field quality measurements to ensure field uniformity and alignment, and validating the magnet's ability to handle thermal and mechanical stresses during operation. Quench protection systems are also tested under real conditions, often by provoking intentional quenches and monitoring voltage growth, energy dissipation, and quench propagation. A prime example of this is the cryogenic qualification campaign of the LHC superconducting magnets at CERN for the cryo-magnets between 2002 and 2006. The dedicated test facility includes 12 fully equipped test benches organized into 6 clusters, each capable of operating at both 1.9 K and 4.5 K. These benches are equipped with high-current power converters and are supported by advanced instrumentation for magnetic and quench diagnostics. During the LHC campaign, over 1,200 dipoles and 400 quadrupoles were tested individually. Each magnet underwent a full sequence of powering, quench training, magnetic measurements, and diagnostic analysis. The extensive testing infrastructure is shown in Fig. 29.

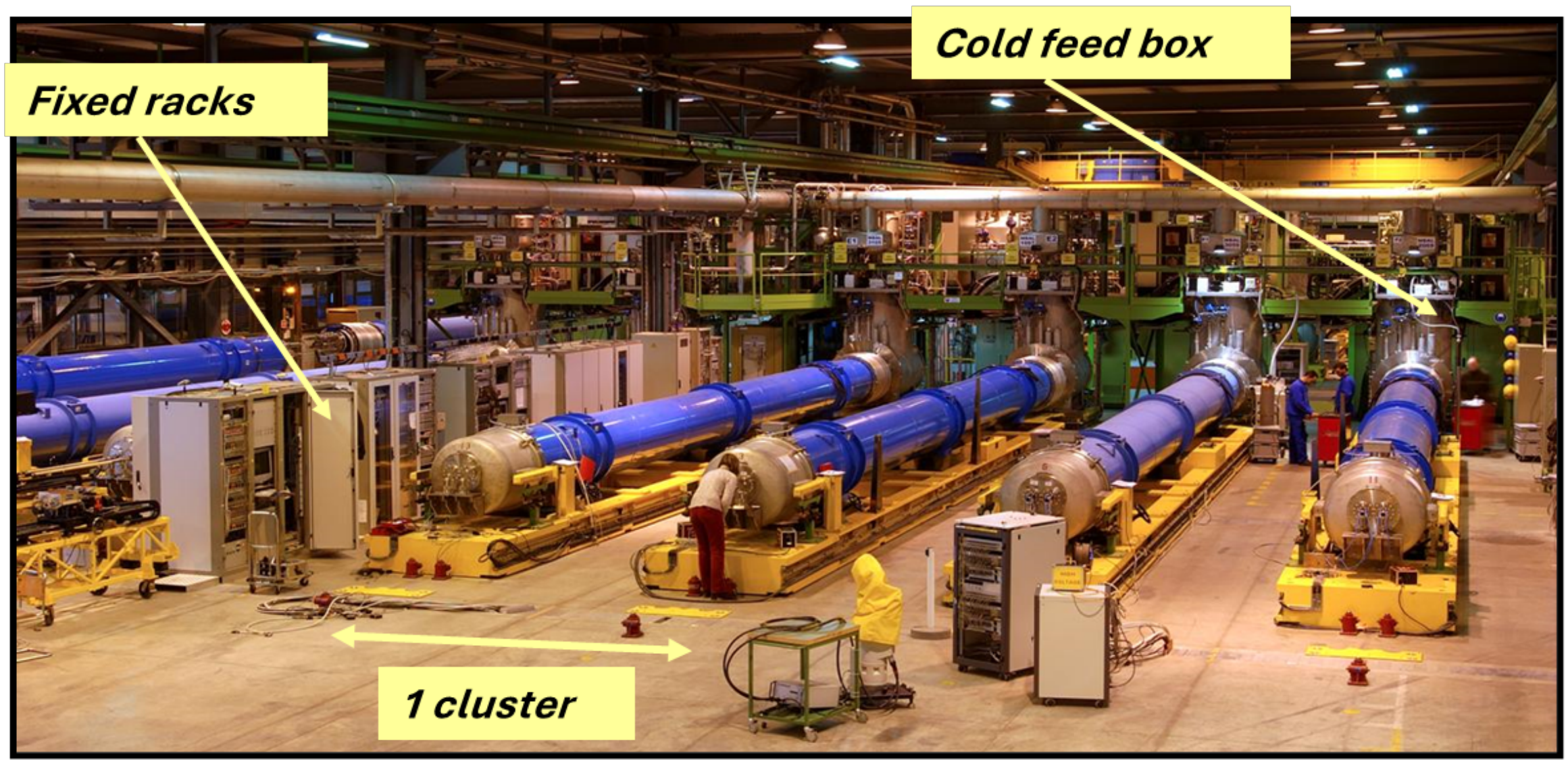

**Fig. 29:** SM18 Hall at CERN, dedicated to the cryogenic qualification of LHC magnets. Magnet performance was tested on 12 benches operating at 4.5 K and 1.9 K. Up to three magnets could be tested simultaneously at cold, while others were in mounting, dismounting, warm-up, or cool-down phases.

Figure 30 shows a standard qualification program for an LHC dipole at 1.9 K and 4.4 K.

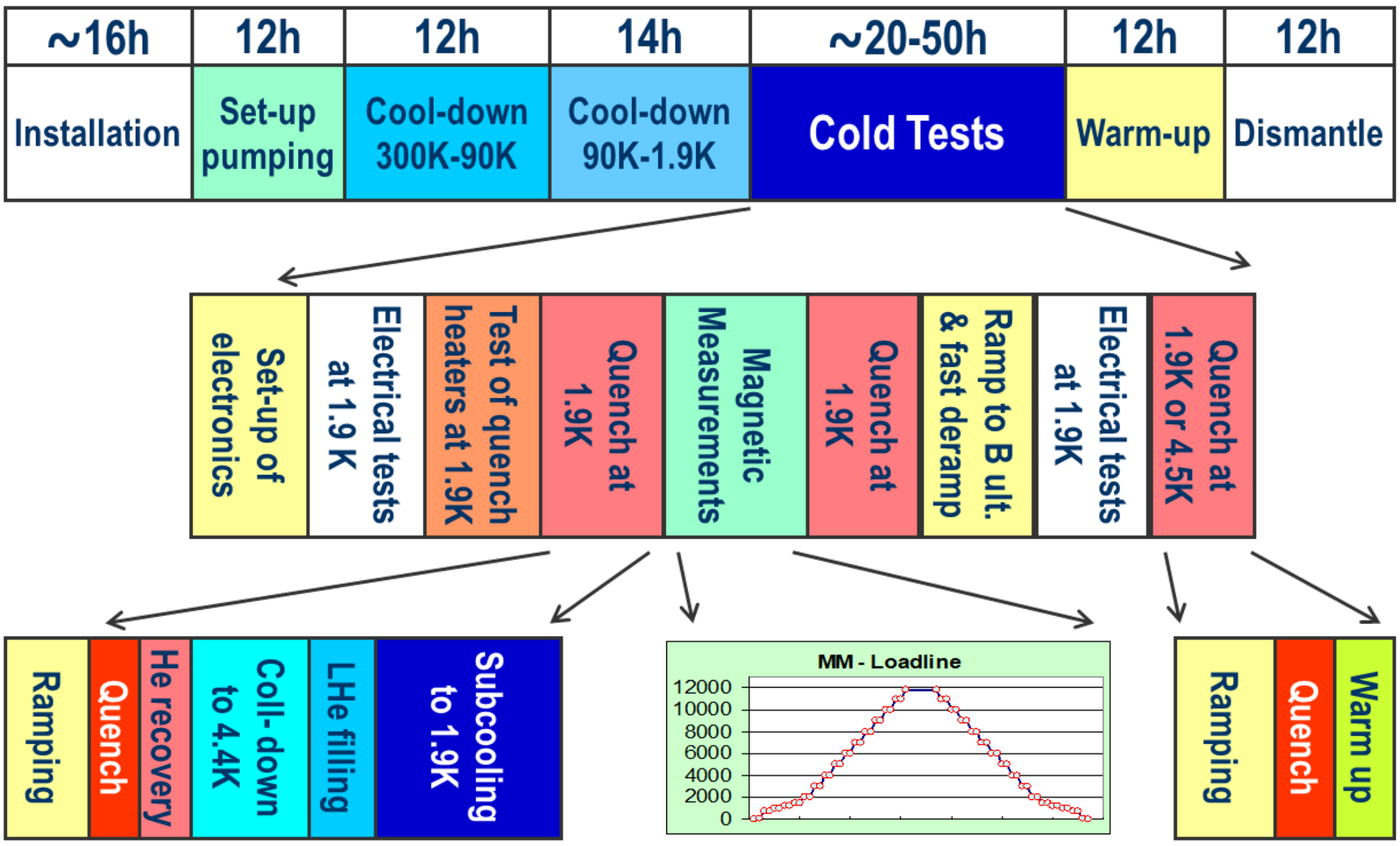

**Fig. 30:** Timeline and sequence of cryogenic qualification tests for LHC superconducting magnets at CERN. The process includes magnet installation, stepwise cool-down from 300 K to 1.9 K, and a comprehensive series of cold tests such as electrical checks, quench validation, and magnetic measurements [21].

The test program included several steps (Fig. 30):

- ***Integrity***
  - The magnet is cooled down from room temperature to 4.5 K and eventually to 1.9 K, depending on the test requirements.
  - Test and acceptance of cryogenic, insulation vacuum, and electric integrity.

- ***Powering and quench protection***
  - The magnet is powered gradually to nominal and ultimate currents.
  - The quench detection and protection system (quench heaters) is activated to detect and safely handle quenches starting with provoked quenches at low current (1.5 kA and 6 kA).
  - Quench diagnostics tools (voltage taps, temperature sensors, quench antennas, fast acquisition systems) monitor key signals to analyze the quench origin and dynamics.
- ***Performance and field quality***
  - Quench training and performance acceptance at 1.9 K: The magnet is subjected to repeated quenches to reach its nominal current capacity.
  - Quenches at 4.4 K to confirm the integrity of the superconducting cables.
  - Ramp rate induced quenches to verify the electromagnetic losses.
  - Memory and stability (for selected magnets): Tests determine if the magnet retains performance after thermal cycles or quenches. Another training quench campaign is carried out after the magnet is warmed up to 300 K and then cooled down again to 1.9 K.
  - Field performance and acceptance at 1.9 K: Magnetic field strength and uniformity are measured (with rotating coils) performed for 20 % of the magnets.
  - Field quality acceptance: Harmonics and field integrals are measured to ensure compliance with design specifications.
  - Dynamic effect studies at 1.9 K (for selected magnets ~10 %) to assess the field quality dependence of the powering history during the injection plateau and the ramp start to nominal field (decay and snap back).
- ***Post-test and magnetic measurements Analysis***
  - Data analysis identifies trends, anomalies, or degradation in performance.
  - Quench location detection uses voltage timing and acoustic sensors to locate the quench initiation point.
  - Field and Harmonic analysis to check whether the field integral and field quality matches with the specifications.

Figure 31 describes typical training curves of LHC magnets. During initial excitations, superconducting accelerator magnets often quench at currents significantly below their target operating levels. These premature quenches are typically caused by intrinsic mechanical disturbances, such as conductor motion or cracking in the epoxy. With repeated excitations, the magnet gradually improves and eventually reaches its nominal performance. This process is known as quench training. More details can be found in [31].

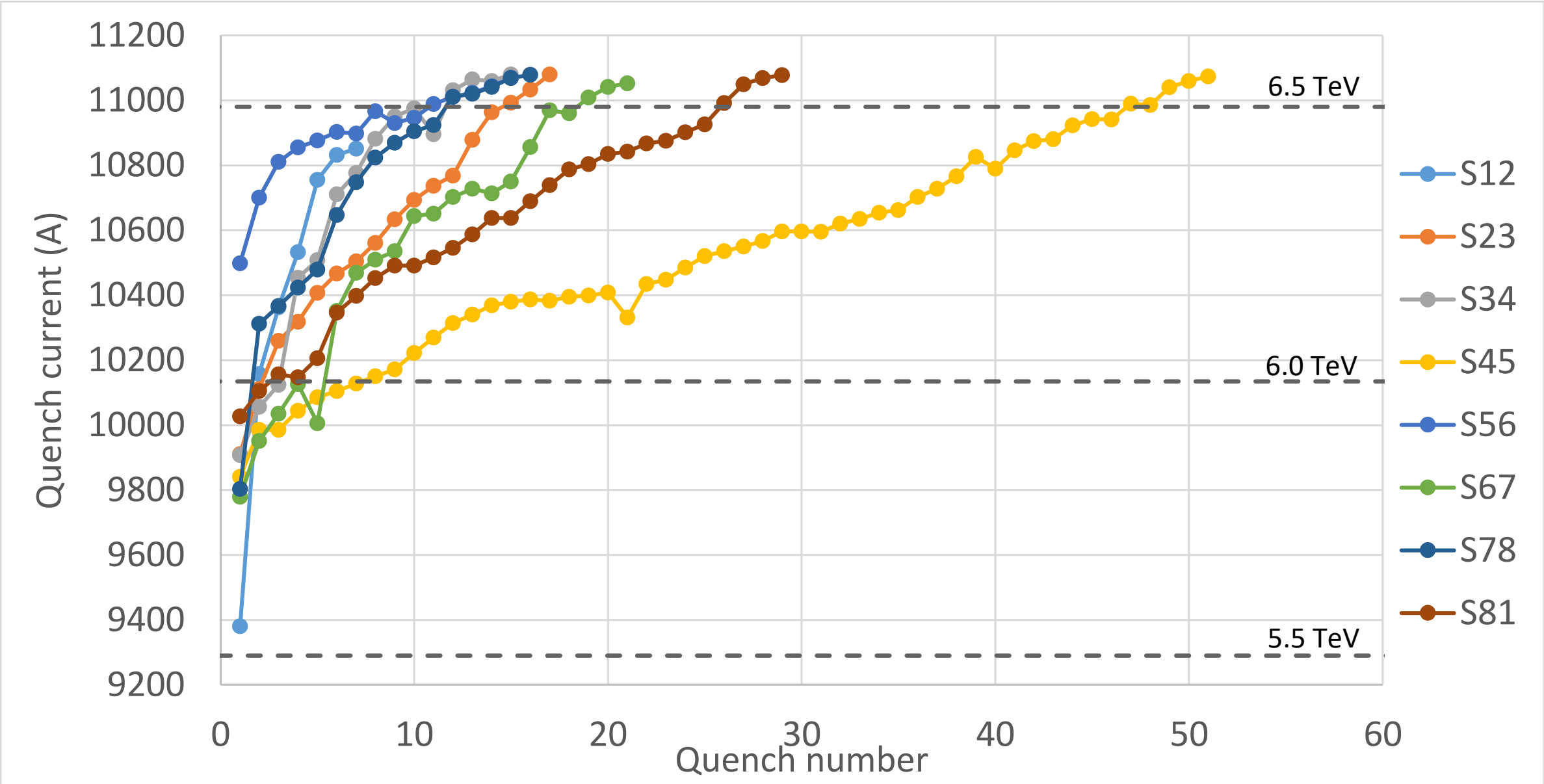

**Fig. 31:** Quench training curves for several LHC superconducting dipoles installed in different sectors.

## 6.2 Dynamic effects at low temperature and magnetic fields: decay and snap back

While static field errors in superconducting accelerator magnets—such as those arising from geometric misalignments or iron saturation—are similar to those found in conventional magnets (see previous course on conventional magnets) and are relatively predictable and reproducible, dynamic field errors are specific to superconducting technology and present more complex challenges. Among these, the most critical ones are the decay and snapback effects occurring at low field during the injection plateau and initial current ramp as observed during powering machine cycles for the LHC. These phenomena are intrinsic to the superconducting state and are particularly pronounced in magnets wound with Rutherford cables composed of many fine filaments. The origin of these dynamic effects lies in the persistent currents induced within the superconducting filaments when exposed to a magnetic field. At low field (e.g., B ≈ 0.54 T), these currents create a screening effect that alters the internal magnetization of the conductor. Due to non-uniform current distribution—caused by variations in inter-strand contact resistance ($R_c$), joints, and cable geometry—this magnetization is not stable and slowly changes with time. As highlighted in Fig. 32, this leads to a decay in the field harmonics, notably the sextupole ($b_3$) and decapole ($b_5$) components, over the long injection flat-top (which can last up to 10’000 seconds). This decay behaves similarly to a rectification process, where a net demagnetization occurs as a result of filament magnetization relaxing toward equilibrium. When the current is ramped up again at the end of injection, the internal magnetization is rapidly re-established due to the increased background field, leading to a snapback—a sudden jump in field components back to their nominal hysteresis path (see Fig. 32). This occurs over a narrow field interval of 15–30 mT and within seconds. The abrupt nature of this change poses significant challenges for beam optics, particularly chromaticity control, since even a 1-unit variation (0.01 %) in $b_3$ can lead to a ±50-unit change in chromaticity—an order of magnitude beyond acceptable tolerances [6].

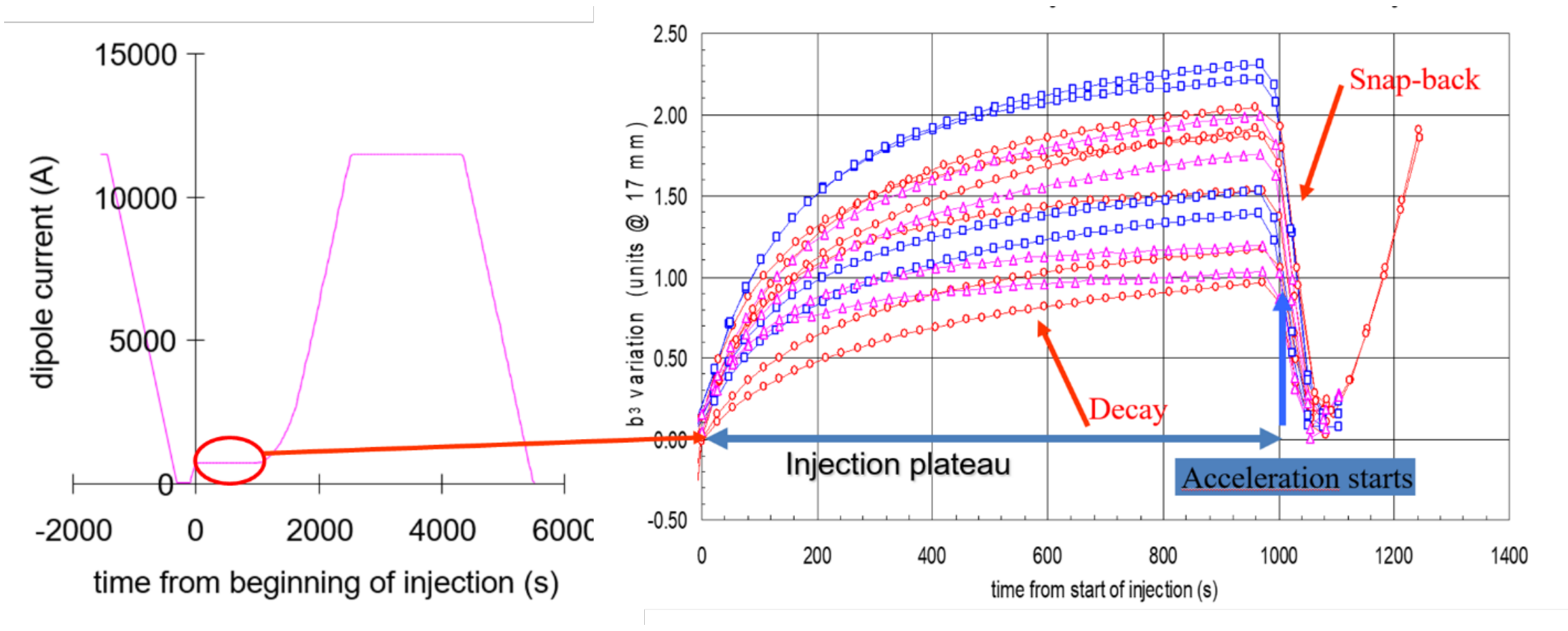

**Fig. 32:** Decay and snap back of the sextupole component $b_3$ measured at 1.9 K in several LHC 15 m long dipole during a current plateau of 760 A (~0.54 T) and 1000 s. The sextupole amplitude changes by some units during the current plateau and bounces back to its initial vales at the beginning of the ramp to nominal field. 2.3 units of sextupole decay corresponds to a change of 115 unit of chromaticity.

Importantly, the amplitude and time constants of decay and snapback are not universal; they depend strongly on the powering history, including previous flat-top current, duration, and the presence of a pre-injection porch. The magnetic history of each magnet affects how persistent currents redistribute within the cable, and these effects are not directly measurable at room temperature. As such, field quality errors due to decay and snapback must be corrected using pre-cycling strategies and look-up tables derived from cryogenic magnetic measurements for each individual magnet.

Moreover, these phenomena highlight the complex interplay between superconducting cable design and magnet performance. For example, changes in strand coatings, cable compaction, or collaring pressure can all influence inter-strand resistance, thereby affecting the magnitude and dynamics of persistent currents. Experimental campaigns at CERN using short dipole models and 15 m long dipoles [15, 25] have shown that magnets with nominally identical designs can exhibit large variations in snapback amplitude and decay behavior, underscoring the need for individual cryogenic testing and correction tuning before magnet installation.In conclusion, dynamic field errors such as decay and snapback are intrinsic to superconducting magnets and must be carefully studied and controlled in high-performance accelerator applications like the LHC. Their impact on field quality—especially $b_3$ and $b_5$ harmonics in dipoles ($b_6$ for the quadrupoles)—requires not only careful magnet design but also precise control of powering procedures and in situ field corrections.

## 7 Conclusion

Throughout this course, we have explored the fundamental physics, material science, and engineering strategies that support the design, construction, and operation of these complex systems. Superconducting magnets enable access to high magnetic fields unattainable with conventional technology, but they introduce stringent requirements in terms of materials, mechanical stability, cryogenics, and quench protection. This course has explored the fundamental principles and applied technologies of superconducting accelerator magnets. Starting with key physical phenomena—zero resistance, the Meissner effect, vortex dynamics, and flux pinning—we examined the distinctions between Type I and II superconductors, and between LTS and HTS materials. Effective vortex pinning and the control of parameters like $T_c$, $J_c$, and $Bc_2$ are essential for stable high-field performance. Quench behavior was a central topic, including its causes, detection, and protection. Systems like quench heaters, dump resistors, and CLIQ illustrate the importance of fast, distributed energy dissipation to protect the magnet during transitions to the normal state. On the engineering side, we reviewed the magnet life cycle—from conductor fabrication to winding, impregnation, and cryostat integration. While NbTi and $Nb_3Sn$ processes are industrially mature, HTS materials such as ReBCO require more specialized handling due to their anisotropy and fragility. Mechanical integrity under strong Lorentz forces is critical. Strategies like pre-stressing, collars, yokes, and innovative stress managed designs were discussed for maintaining coil stability and minimizing training. Case studies, including the LHC dipoles, conduction cooled SLS2.0 superbends, Non-Insulated HTS tapes for high field capture solenoids, and superferric dipoles, highlighted the breadth of superconducting magnet applications and the tailored solutions each requires. Cryogenic testing, quench training, and field validation—supported by dedicated infrastructures like CERN's SM18—are crucial to qualify superconducting magnets before installation. Emphasis must also be placed on understanding and mitigating dynamic, powering-history-dependent field errors—such as decay and snapback of the magnetization—which can significantly affect beam stability and machine performance.

The outlook for superconducting accelerator magnets remains extremely promising. As particle accelerators push the limits of beam intensity, luminosity, and energy, the demand for high-field, compact, and stable magnets will only increase. In parallel, the need for more energy-efficient and sustainable magnet technologies is becoming a key priority, both to reduce operational costs and to align with broader environmental goals. The continued development of HTS materials is expected to open new frontiers in magnet design, enabling not only higher performance but also improved efficiency and reduced infrastructure demands.

## Acknowledgements

The author gratefully acknowledges the support of the Magnet Section at the Paul Scherrer Institute, with special thanks to Rebecca Riccioli, Ciro Calzolaio for their contributions to the preparation of this lecture and Carolin Zoller and Quentin Gorit for correcting the manuscript. Special thanks to my colleagues A. Ballarino, L. Bottura, M. Buzio, G. De Rijk, P. Ferracin, P. Pugnat, C. Senatore, A. Siemko, and E. Todesco for providing information, material and support during many years.